\documentclass[review]{elsarticle}

\usepackage{lineno,hyperref}

\usepackage{graphicx}
\usepackage{subfig}
\usepackage{float}
\usepackage{setspace}
\usepackage{array}
\usepackage{multicol}

\journal{Journal of \LaTeX\ Templates}

\begin{document}

\begin{frontmatter}

\title{A high rate and high timing photoelectric detector prototype with RPC structure}

\author[address1,address2]{Yiding Zhao\fnref{1}}

\author[address1,address2]{D.Hu\corref{correspondingauthor}\fnref{1}}
\cortext[correspondingauthor]{Corresponding author}
\ead{hurongr@ustc.edu.cn}

\author[address1,address2]{M.Shao\corref{correspondingauthor2}}
\cortext[correspondingauthor2]{Corresponding author}
\ead{swing@ustc.edu.cn}

\author[address1,address2]{Y.Zhou}
\author[address1,address2]{S.Lv}
\author[address1,address2]{Xiangqi Tian}
\author[address1,address2]{Anqi Wang}
\author[address1,address2]{Xueshen Lin}
\author[address1,address2]{Hao Pang}

\author[address1,address2]{Y.Sun}

\address[address1]{State Key Laboratory of Particle Detection and Electronics, University of Science and Technology of China, 96 Jinzhai Road, Hefei 230026, China}
\address[address2]{Department of Modern Physics, University of Science and Technology of China (USTC), 96 Jinzhai Road, Hefei 230026, China}
\fntext[1]{These authors contributed equally to this work}

\begin{abstract}
To meet the needs of high counting rate and high time resolution in future high energy physics experiments, a prototype of gas photodetector with RPC structure was developed. Garfield++ simulated the detector's performance, and the single photoelectron performance of different mixed gases was tested with an ultraviolet laser. The detector uses a low resistivity ($\sim1.4\cdot 10^{10} \Omega\cdot cm$) float glass so that its rate capability is significantly higher than that of ordinary float glass($10^{12}\sim10^{14} \Omega\cdot cm$), the laser test results show that in MRPC gas($R134a/iC_{4}H_{10}/SF_{6}(85/10/5)$), the single photoelectron time resolution is best to reach 20.3 ps at a gas gain of $7\cdot 10^{6}$. Increasing the proportion of $iC_{4}H_{10}$ can effectively reduce the probability of photon feedback, without changing the time resolution and maximum gain. In addition to being applied to high-precision time measurement scenarios (eg:T0, TOF), the detector can also quantitatively test the single photoelectron performance of different gases and will be used to find eco-friendly MRPC gases.
\end{abstract}


\end{frontmatter}



\section{Introduction}
\paragraph{}
With the development of particle and nuclear physics research towards high energy and high luminosity in the future, new challenges have been posed to the performance of particle detectors, such as time resolution and counting rate. Resistive plate chambers (RPCs)\cite{firstRPC1,firstRPC2} are commonly used in accelerator and non-accelerator based experiments and are responsible for triggering\cite{triggerRPCinATLAS}, timing\cite{timingRPC,timingRPCforTOF} etc, its close relative in technology, multigap RPC (MRPC), is often used to measure the time of flight(TOF) of particles to
achieve particle identification\cite{firstMRPC,MRPCTOF}. In current nuclear and particle physics experiments, the timing accuracy of the commonly used MRPC is generally not better than 50 ps, and because the resistivity of the resistive material used (float glass or bakelite ) is generally in the order of $10^{12}\sim 10^{14}\Omega\cdot cm$, the rate capability is limited\cite{rateofRPC,HighRateModeofRPC,LowRMRPC}. Future experiments on the measurement of TOF is bound to put forward higher requirements. Such as SoLID, in order to achieve 3 sigma separation of more than 7 GeV pi-K, the accuracy requirements of TOF measurement is about 20 ps accuracy requirements\cite{Solid20ps}. In short, the development of high time resolution, high counting rate, large area, low-cost detector, is driven by the experimental needs.

Photosensitive gaseous detectors have advantages over other types of photodetectors mainly in terms of size and
cost\cite{photoE_applyinGasdetector}. However, the sensitivity is mainly limited to ultraviolet (UV) photons, as only photocathodes that are not sensitive to gases or certain photosensitive gas mixtures can be used\cite{pad-photonDetector-Cherenkev}. Nevertheless, the UV sensitivity is suitable for Cherenkov and UV-scintillation detectors, and photo-sensitive gaseous detectors with cesium iodide (CsI) photocathodes have been widely used\cite{PhotoGasDecAndApplication}. With a planar uniform geometry and atmospheric pressure of the gas, they can be easily enlarged without diminishing the time resolution. A time resolution of 44 ps for a single photoelectron was demonstrated by a PICOSEC-Micromegas detector. As is the case for this detector, micro-pattern gaseous detectors (MPGDs) are used for the photoelectron multiplication in order to suppress photon and ion feedback to the photocathode\cite{PICOSEC}. Utilizing parallel-plate avalanche chambers\cite{PPCwithCsI} and resistive plate chambers\cite{photoRPCinRICH,hignRatePositionPhotoRPC,RPCwithCsI} were also investigated. Although the photon and ion feedback problems are more serious, they can outperform PICOSEC in terms of optical detection efficiency and time resolution because the higher electric field near the photocathode enhances the quantum efficiency, and the uniform high electric field enables the initial photoelectrons to rapidly multiply and reduce diffusion, therefore, the time resolution is better. Besides, the simple structure can reduce the cost.\cite{photoRPC25ps}

This paper focuses on the needs of future high energy physics experiments, and develops a prototype of gaseous photodetector with high time resolution (single photoelectron, 20 $\sim$ 30 ps), high rate capability with RPC structure. The detailed information is show in this paper.

\section{Detector Design}

\begin{figure}
    \centering
    \includegraphics[width=1\linewidth]{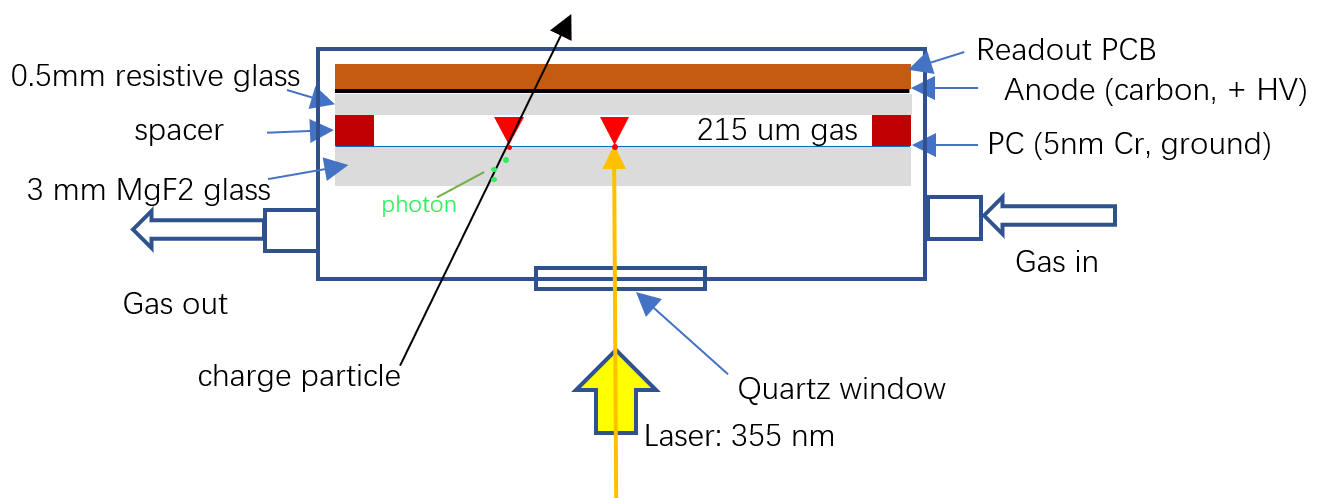}
    \caption{Schematic design of detector.}
    \label{fig:RPC-schematic}
\end{figure}

The diagram of the detector is shown in Fig. \ref{fig:RPC-schematic}. The whole detector is fixed in a closed stainless steel chamber with ventilation. Detector has a RPC structure, the main change is one electrode was replaced by a chromium (Cr) photocathode, so that the detector has the ability of photon detection. From bottom to top, there are: quartz window, photocathode(MgF2 base, 5 nm Cr plated on one surface), 215 $\mu$m gas gap, resistive plate(float glass with different resistivity), readout PCB (application of high voltage and inducing signal).

\begin{figure}
    \centering
    \includegraphics[width=1\linewidth]{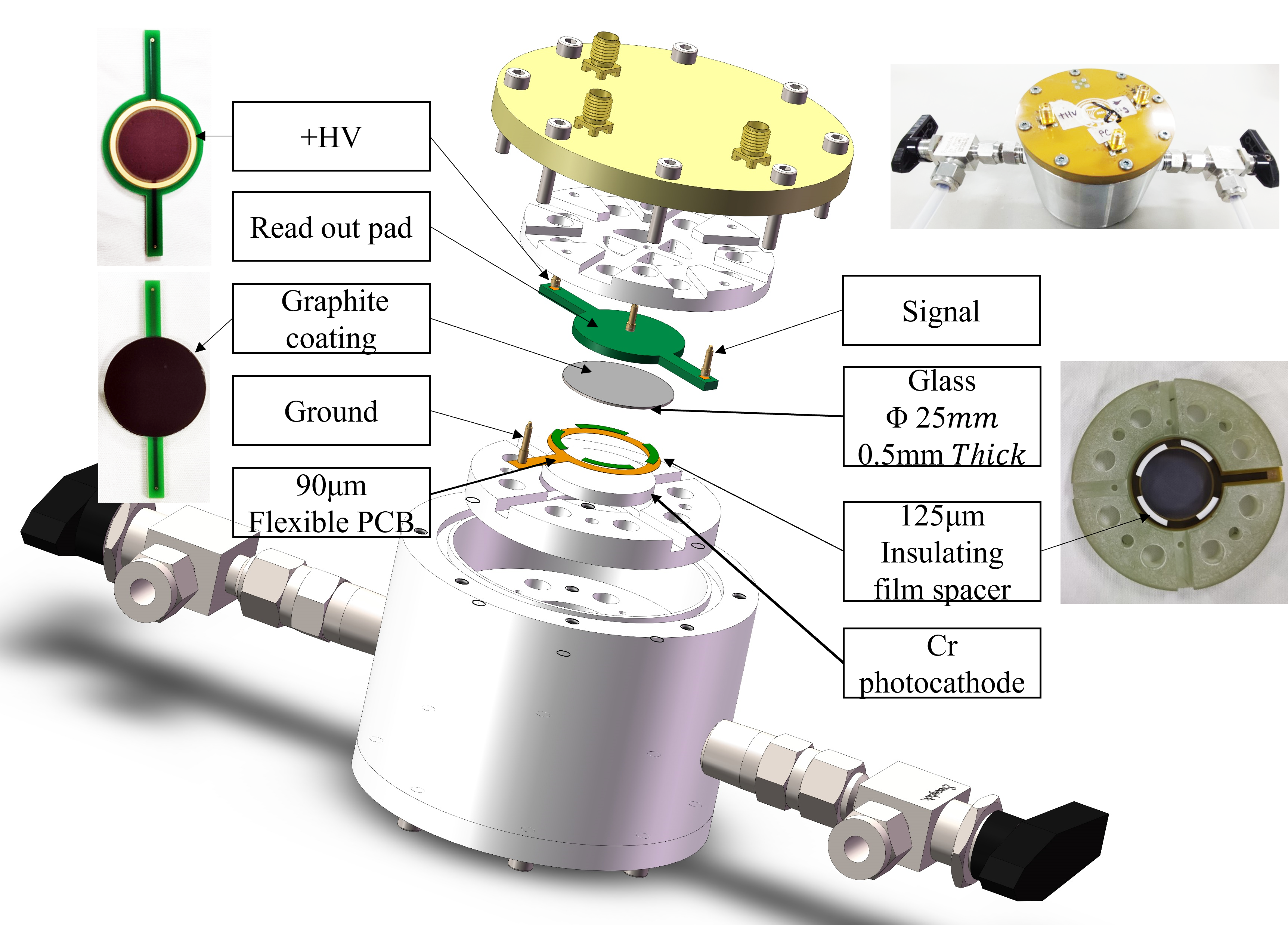}
    \caption{Exploded view and photographs of the detector.}
    \label{fig:install}
\end{figure}

The detector is simple in structure and can be installed in the air. According to the design, each module of the detector was manufactured and assembled in the ultra-clean room, as shown in Fig. \ref{fig:install}.

\section{Simulation}

ANSYS was used to simulate the electric field and weighting field distribution of the detector, Garfield++ was used to simulate the parameters of different gases, the gain and time resolution of single electron.

\subsection{ANSYS simulation}
The detector was modeled in ANSYS, and the electric field and weighting field are simulated. 

\begin{figure}
    \centering
    \includegraphics[width=1\linewidth]{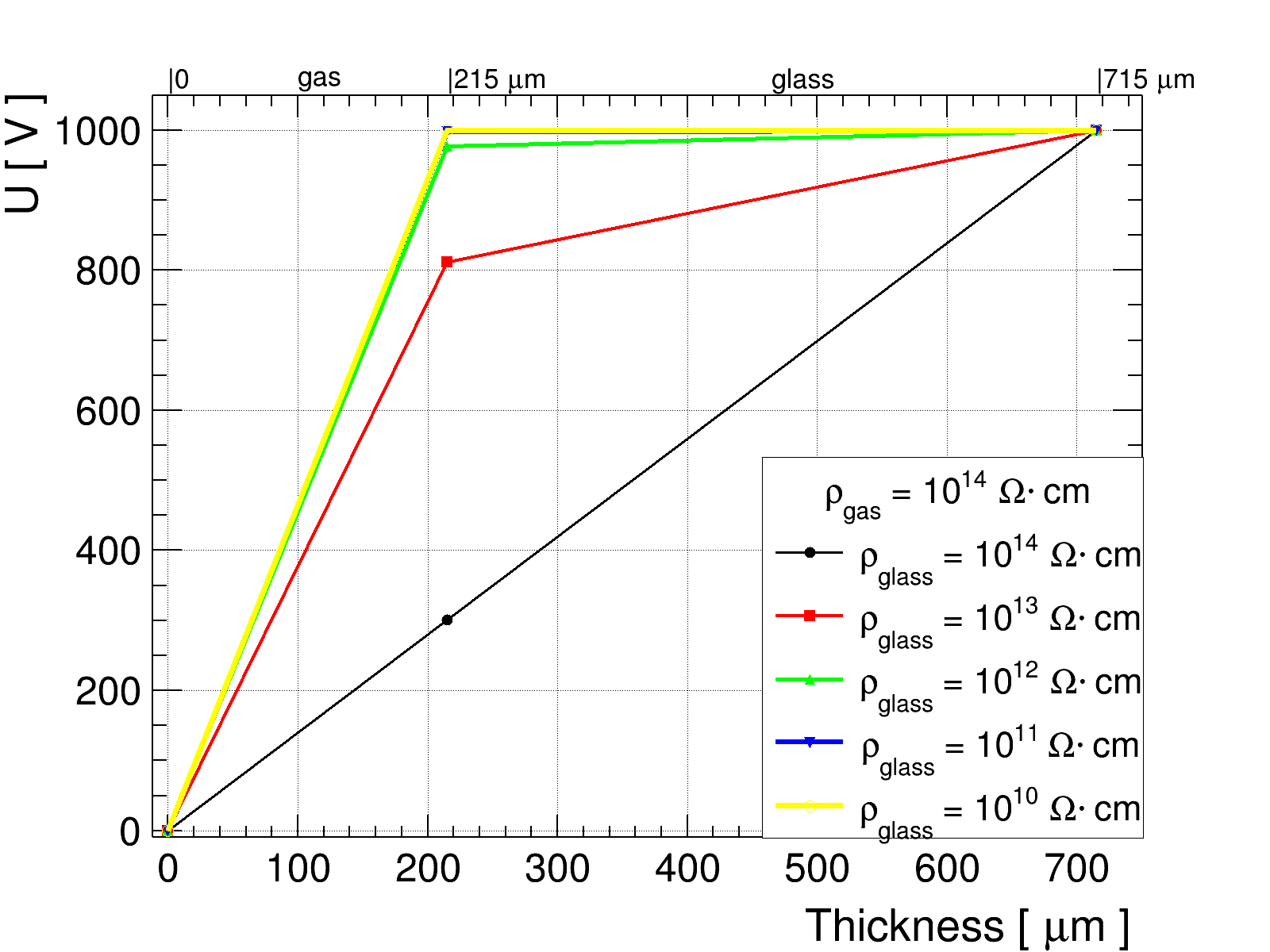}
    \caption{Potential distribution of gas gap with different resistivity glass $\rho_{glass}$.}
    \label{fig:Potential-distribution}
\end{figure}

Due to the existence of a small leakage current in the whole loop, the electric field distribution in RPC should consider the influence of current. In ANSYS simulation, a physical field model containing current was selected to calculate the potential distribution, the anode potential is set to HV = +1000 V and the cathode is grounded. The resistivity of the gas is set to $\rho_{glass}$ = $10^{14}$ $\Omega\cdot cm$, the potential distribution with different resistivity glass was calculated, and the result is shown in Fig. \ref{fig:Potential-distribution}.

For a closed loop containing a current, according to Ohm’s law and resistivity expression $R = \rho \frac{l}{s}$. The potential in the gas gap can be obtained as:
\begin{equation}
    \Delta U_{gas} = HV\cdot\frac{\rho_{gas}\cdot l_{gas}}{\rho_{gas}\cdot l_{gas}+\rho_{glass}\cdot l_{glass}}
\end{equation}

where $\Delta U_{gas}$ is the potential difference in gas gap, HV is the applied voltage, $\rho$ and $l$ are the resistivity and thickness of glass and gas gap. The simulation results agree with the potential distribution obtained by Ohm’s law. When the applied voltage and gas resistivity are fixed, the lower the resistivity of the resistive glass, the greater the potential difference and the electric field in the gas gap.

The distribution of the weighting field (electrostatic) was also calculated. According to Ramo's theorem\cite{Raether1964ElectronAA,induceSignalofRPC}, the potential of the readout electrode is set to 1 V, the photocathode is grounded, the parts between the photocathode and the readout electrode are: gas (215 $\mu m$, $\epsilon$ = 1), float glass(500 $\mu m$, $\epsilon$ = 4.2), graphite(100 $\mu m$, $\epsilon$ = 1), PCB board(120 $\mu m$ FR4, $\epsilon$ = 4.4). The simulation results indicate that the weighting field is $\Delta U \sim$ 0.43 V in the gas gap.

\subsection{Garfield ++ simulation }

The Magboltz\cite{MagboltzMC,magboltzWeb} program in Garfield++\cite{garfieldWeb} was used to calculate macroscopic parameter (townsend coefficient, electron drift velocity) of gas. The single photoelectron time resolution was roughly calculated by the formula\cite{RPC2003}:

\begin{equation}
    \sigma_{t} = \frac{1.28}{(\alpha - \eta)\cdot v}
\end{equation}

\begin{figure}
    \centering
    \includegraphics[width=1\linewidth]{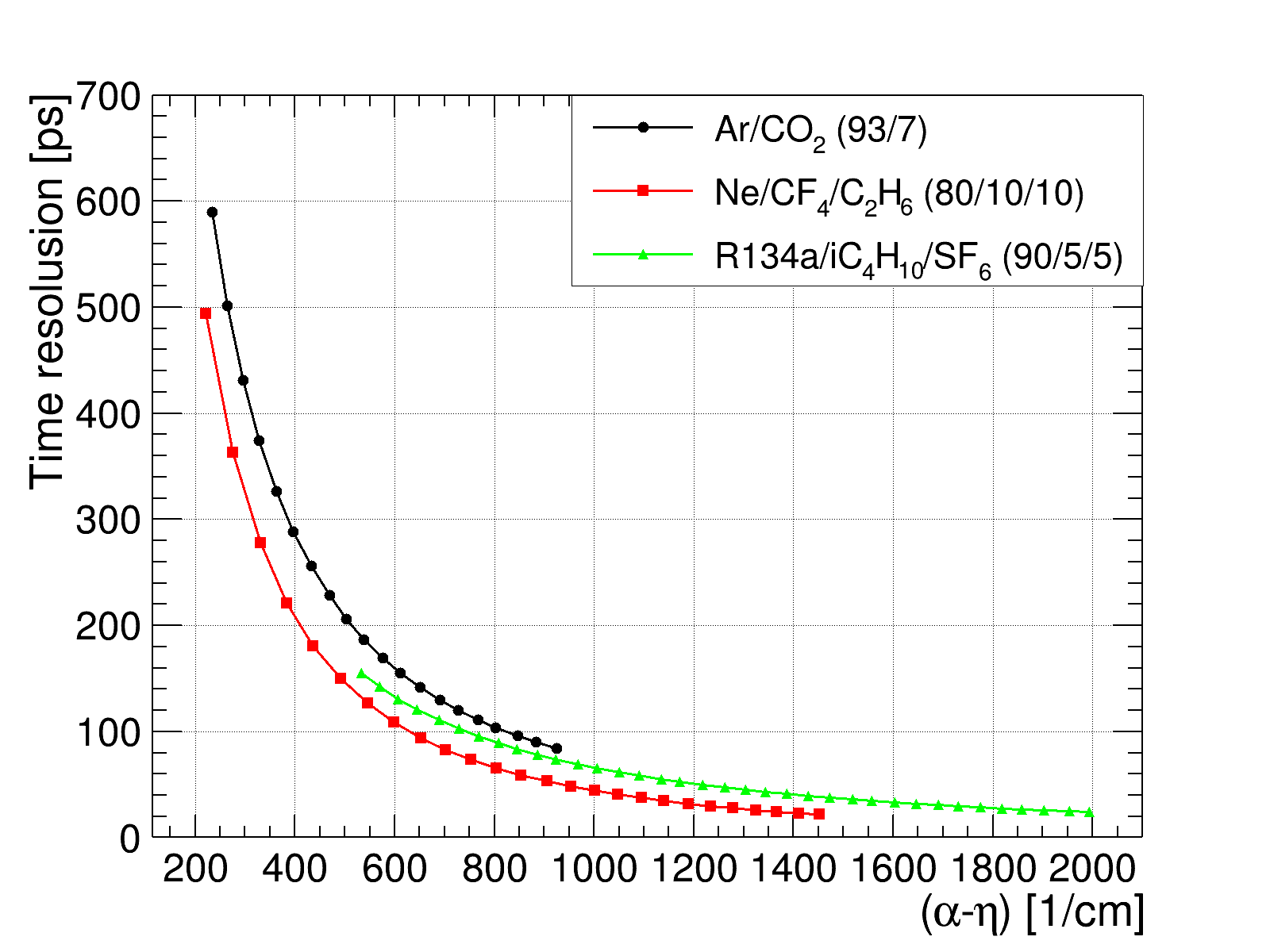}
    \caption{Time resolution vs. Effective Townsend coefficient ($\alpha - \eta$).}
    \label{fig:GasSim}
\end{figure}

The results are shown in Fig. \ref{fig:GasSim}. The time resolution at the same gain: COMPASS gas $<$ MRPC gas $<$ Ar, but since the detector in MRPC gas can work at very high voltage, it is speculated that optimal time resolution can be achieved in MRPC gas.

\begin{figure}
    \centering
    \includegraphics[width=1\linewidth]{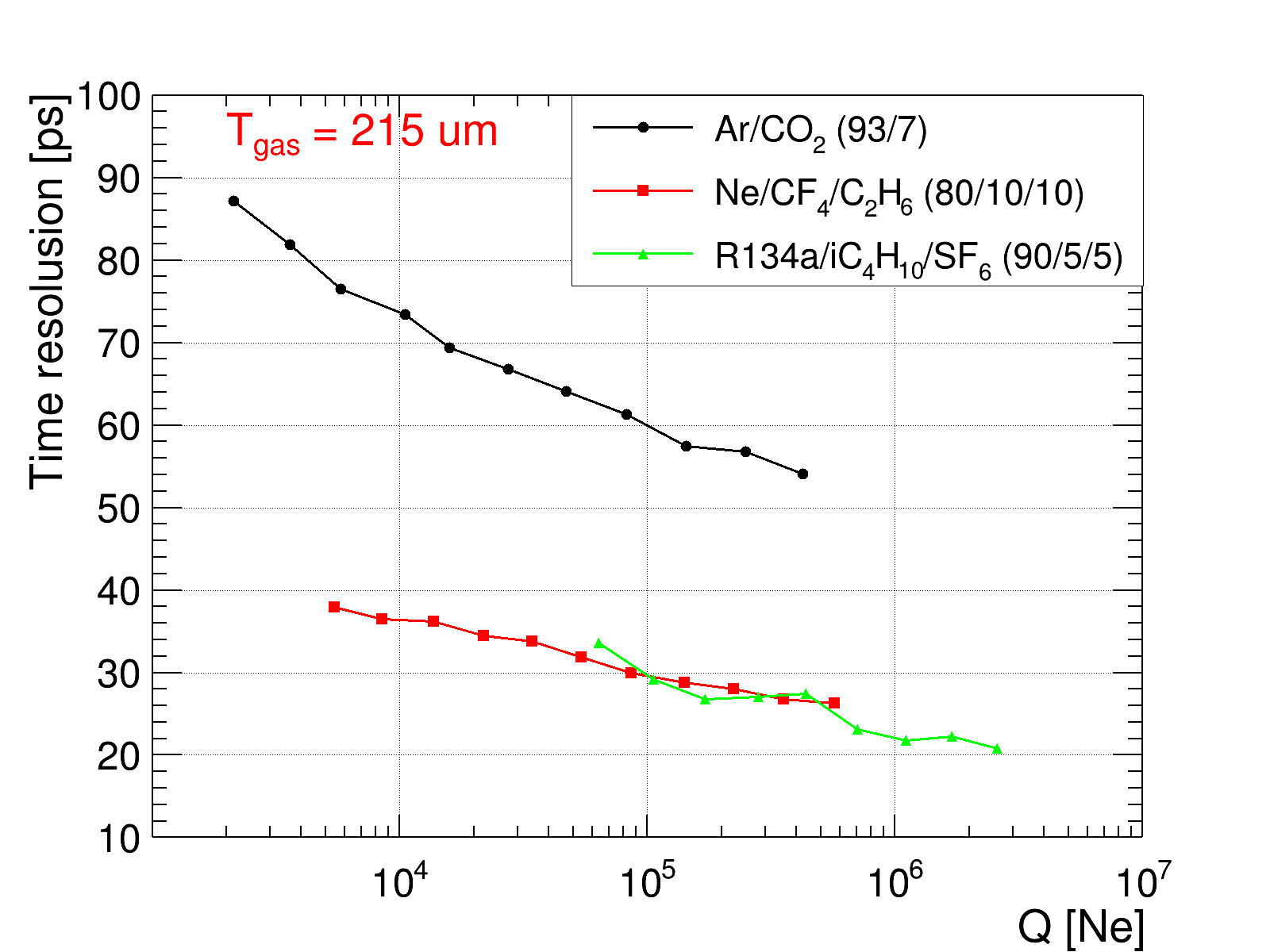}
    \caption{Time resolution vs. induced signal charge (Q) in Garfield++ simulation.}
    \label{fig:GarfieldSim}
\end{figure}

In Garfield++, the avalanche process can be simulated and the corresponding induction signal can be obtained through the weighting field\cite{induceSignalofRPC}, so we can obtain the time resolution of a single photoelectron at different gain. The simulation parameters are the same as the experimental device, the simulation results are shown in Fig. \ref{fig:GarfieldSim}, the time resolution in $Ar/CO_{2}$ is the worst (50 $\sim$ 90 ps), COMPASS gas and MRPC gas have similar time resolution (20 $\sim$ 40 ps) at the same gain.

\section{Experimence and results}

\subsection{Test system}

\begin{figure}[htbp]
\centering
\subfloat[]{%
    \includegraphics[width=0.4\linewidth]{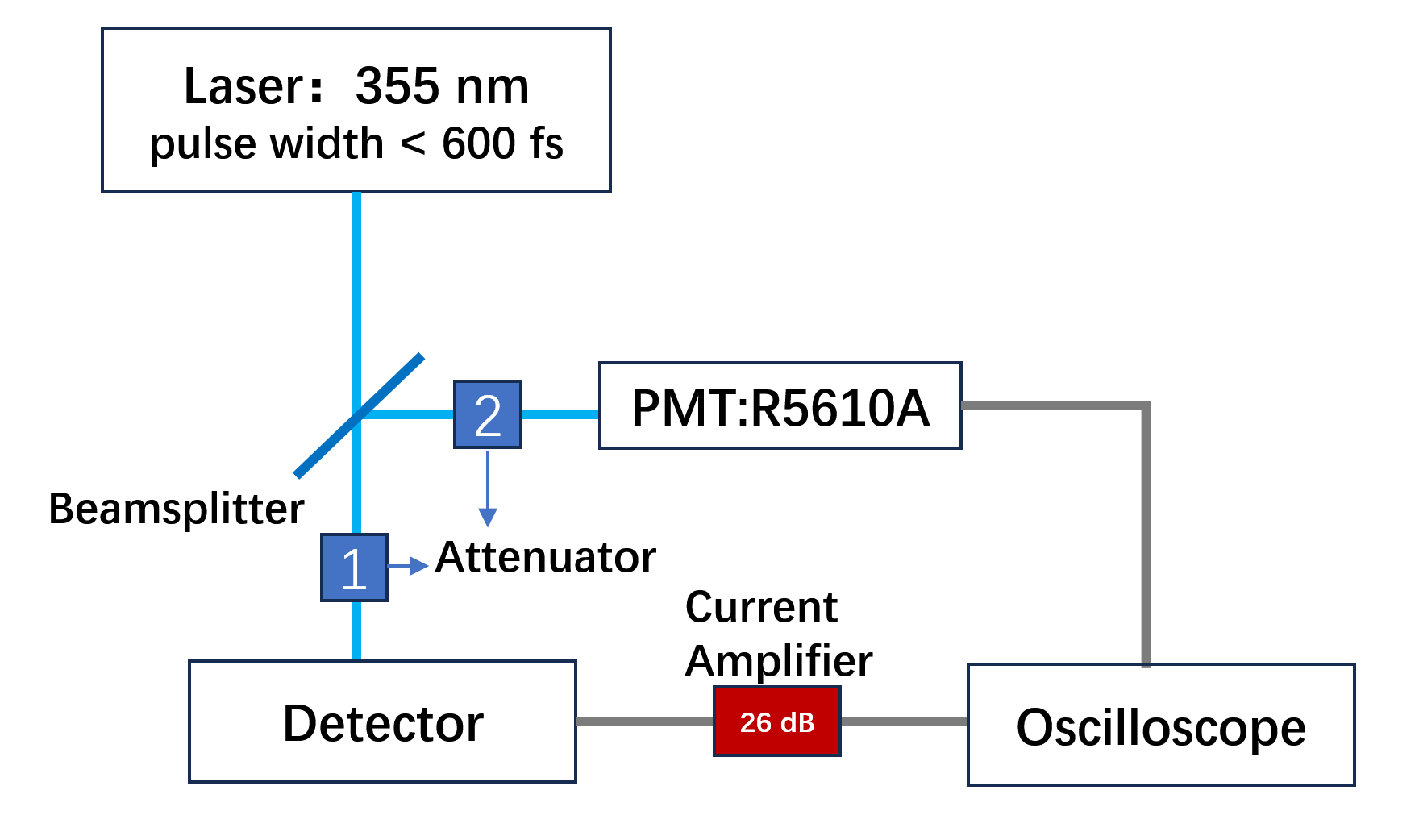}
    \label{fig:subfig_System_diagram2}
}
\hfill
\subfloat[]{%
    \includegraphics[width=0.4\linewidth]{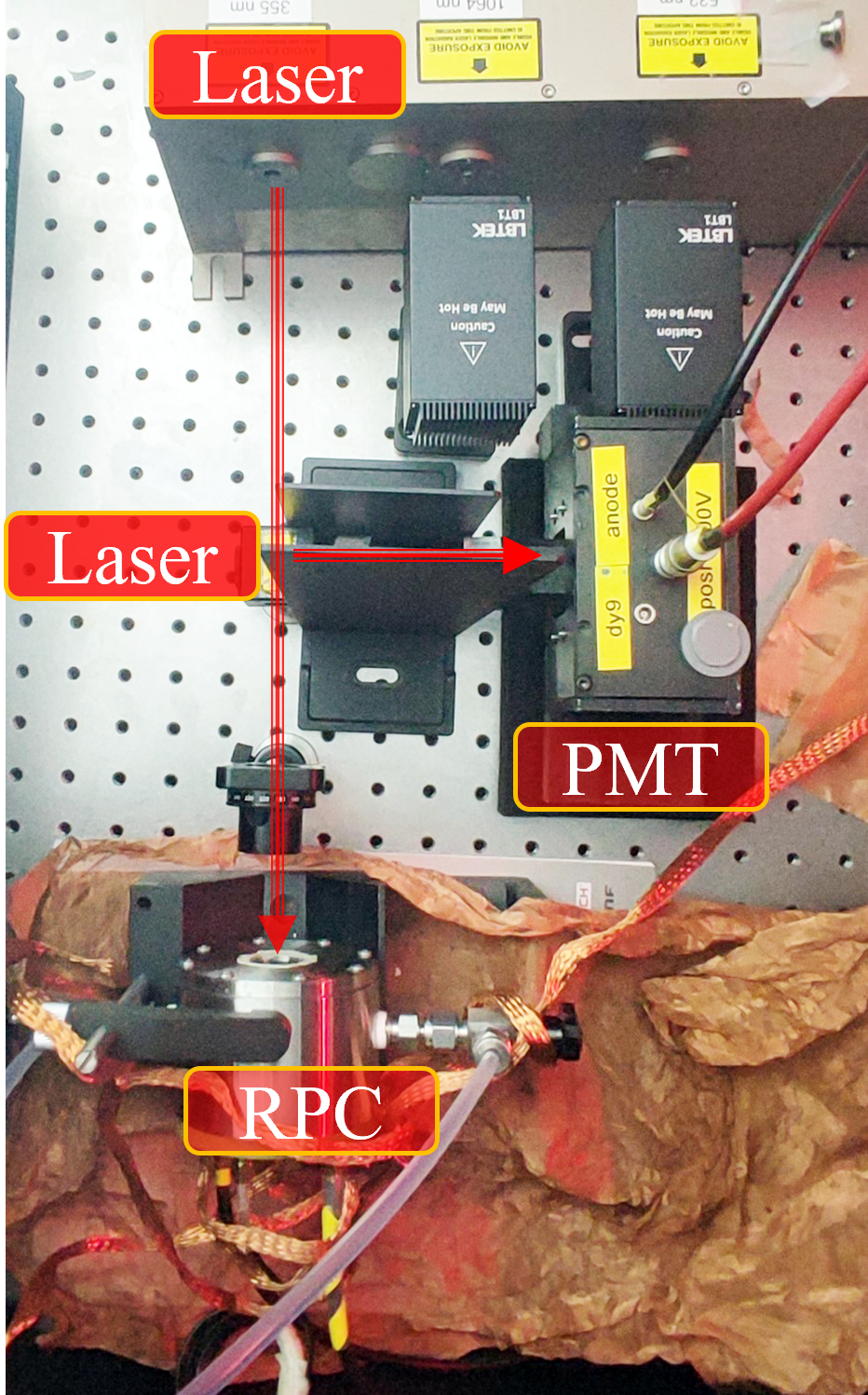}
    \label{fig:subfig_TestSystem2}
}
\caption{Test system. (a) Schematic. (b) Photograph of test system.}
\label{fig:TestSystem}
\end{figure}

The test system was built on the optical platform, as shown in Fig. \ref{fig:TestSystem}. The laser beam is divided into two parts by the beamsplitter, one enters the reference timing PMT(R5610A), and the other enters the detector after passing through the optical attenuator. The two signals of RPC and PMT are collected by the oscilloscope.


The laser is the AOPico laser system produced by Advanced Opowave Corporation. The energy and frequency of the pulse can be adjusted. The wavelength is 375 nm, the pulse width is $<$600 fs. The two signals are displayed on the oscilloscope with an input impedance of 50 $\Omega$. In the pre-test, the frequency of the output laser pulse was set to 4 Hz, the voltage of PMT was set to 280 V, and the amplitude of the PMT signal was controlled in the range of 45 $\sim$ 60 mV by adjusting the attenuator 2. Under this condition, the timing accuracy of PMT is better than 10 ps, which meets the test requirements. The working gas of RPC is $Ar/CO_{2}(93/3)$, adjust attenuator 1 so that the RPC has a significant signal (the intensity of the incident laser determines the number of photoelectrons produced by the photocathode). The time resolution is obtained by the constant-fraction discrimination(CFD). After calibration, the time resolution of the system is about 10 ps (see section 4.3).

\subsection{Choice of resistive plate}

\begin{figure}
    \centering
    \includegraphics[width=1\linewidth]{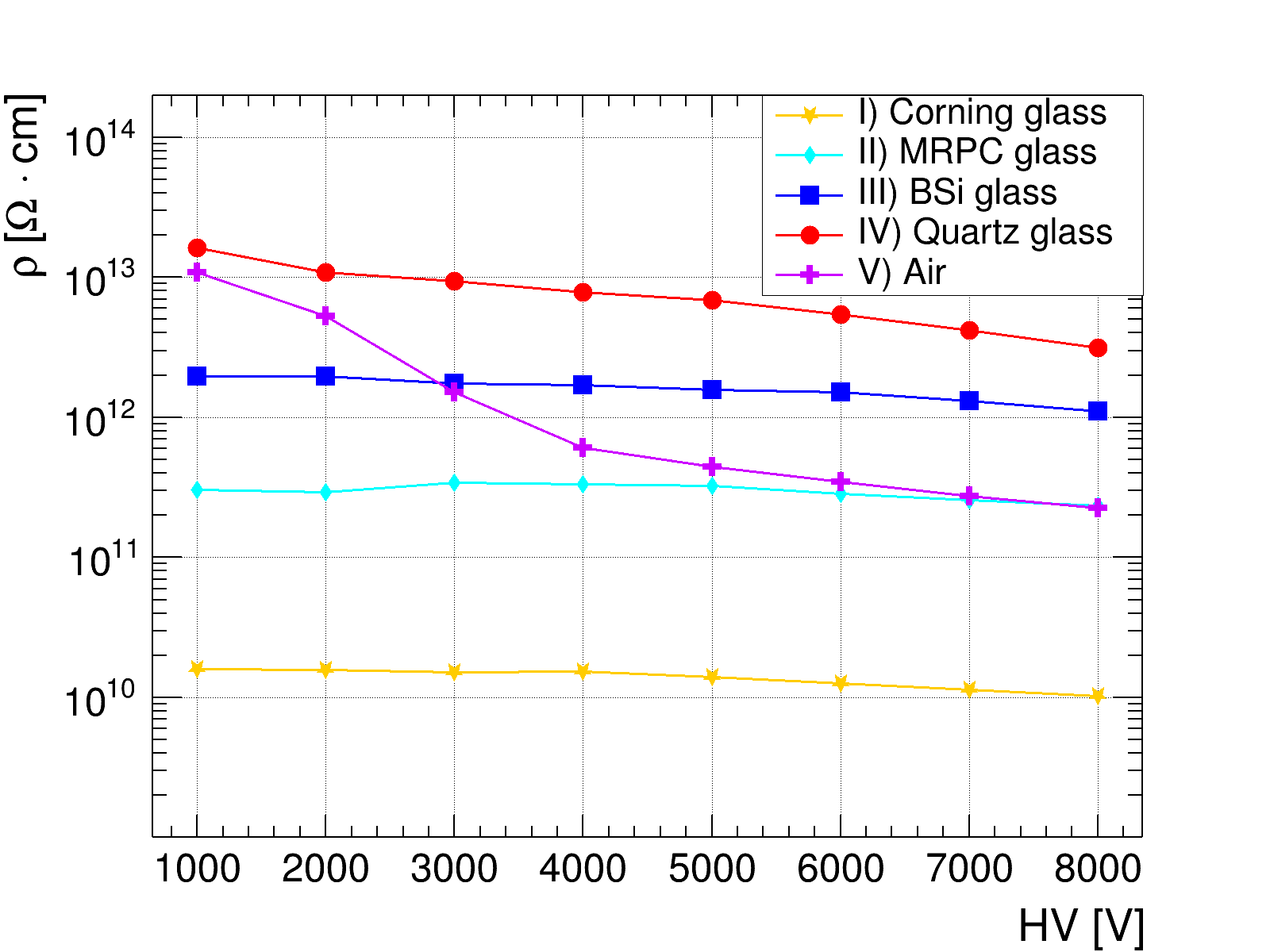}
    \caption{Resistivity test results of glass. The resistivity of air (without glass placed) was also tested.}
    \label{fig:GlassResitivity}
\end{figure}

For resistive gaseous detectors, the resistive plate material (thickness and resistivity) is particularly important for the detector's rate capability. In cooperation with a glass factory, we customized a batch of float glass made of different materials with a diameter of 25 mm, and carried out the corresponding resistivity test\cite{BookRGD_testResitivity}. The results are shown in Fig. \ref{fig:GlassResitivity}. As the voltage increases, the resistivity of the glass will decrease slightly. The resistivity of the air is also tested (no glass was placed, only the 0.5 mm gas gap was left). It is believed that the resistivity of the working gas should be larger than air($10^{13} \Omega \cdot cm$). 


\begin{figure}[htbp]
\centering
\subfloat[]{%
    \includegraphics[width=0.5\linewidth]{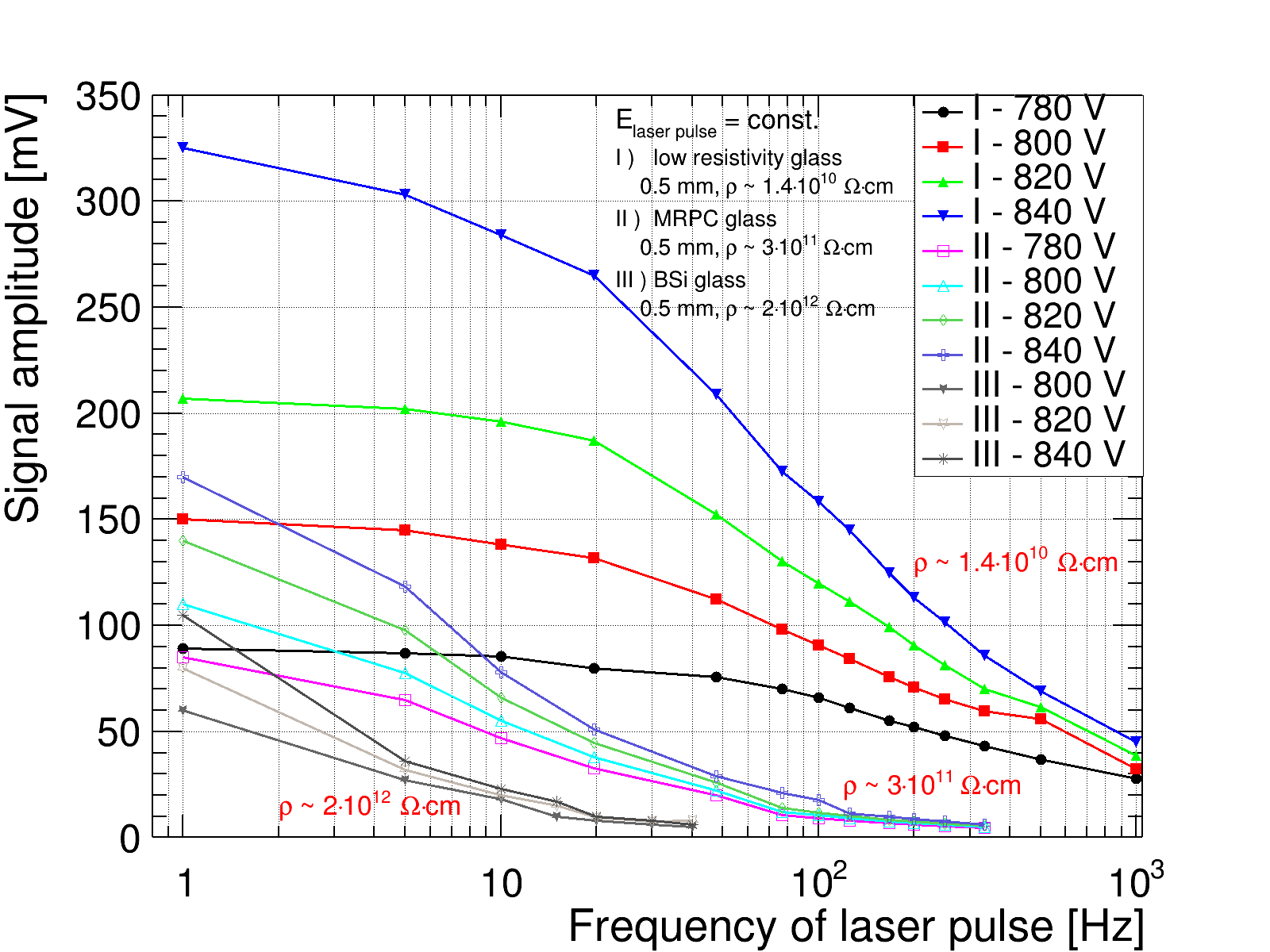}
    \label{fig:subfig-RealA}
}
\subfloat[]{%
    \includegraphics[width=0.5\linewidth]{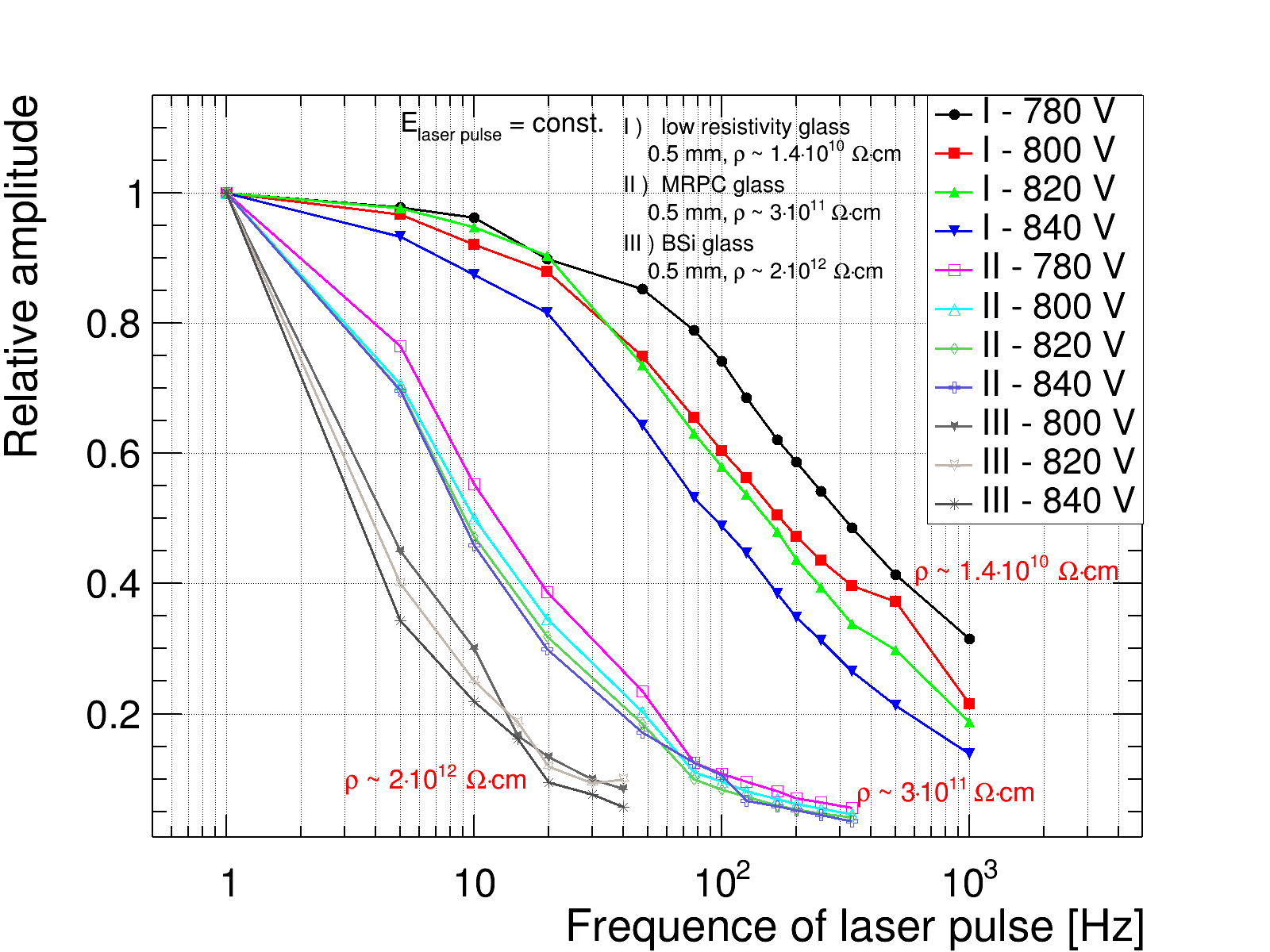}
    \label{fig:subfig-RaletiveA}
}
\caption{The variation of RPC's signal amplitude with laser frequency (counting rate). The laser frequency varies from 1 Hz to 1000 Hz, keeping the energy of the incident laser pulse constant, and the working gas is $Ar/CO_{2}(93/7)$. (a) Signal amplitude collected by oscilloscope. (b) Relative amplitude (maximum amplitude is 1).}
\label{fig:RateTest}
\end{figure}

The detector was assembled with four types of glass and tested with a laser. The energy of the single laser pulse incident on the photocathode was kept constant so that the number of photoelectrons generated by the photocathode was kept the same. $Ar/CO_{2}(93/7)$ was used as the working gas, and the amplitude of the signal varied with the frequency of the incident laser. The results are shown in Fig. \ref{fig:RateTest}.


As can be seen in Fig. \ref{fig:subfig-RealA}, under the same voltage and the same frequency, the higher the glass resistivity, the smaller the signal amplitude, which means that the avalanche size in the gas gap is smaller and the electric field in the gas gap is weaker. This is consistent with the electric field simulation in Section 3.1: Under the same applied voltage, the greater the glass resistivity, the smaller the gas gap potential difference (smaller electric field), resulting in a smaller gain. Fig. \ref{fig:subfig-RaletiveA} shows that as the frequency of the incident laser increases (the counting rate increases), the signal amplitude decreases rapidly, and the higher the glass resistivity, the faster the decay. It should be noted that when quartz glass($10^{13}$ $\Omega\cdot cm$) was used, no significant signal was observed, even at the laser frequency of 1 Hz, it may be because the potential difference in the gas gap is too small due to large resistivity of quartz glass \ref{fig:Potential-distribution}, the electric field in gas gap is not big enough for an avalanche.

The signal weakens with an increase in the counting rate, which is a characteristic of a resistive gaseous detector. When RPC is working, the accumulated charge on the resistive plate will lead to the local electric field drop, and the speed of the accumulated charges release will affect the electric field of the gas gap in the next avalanche, thus affecting gain. The release speed of the accumulated charge on the resistive plate can be characterized by the time constant $\tau$\cite{induceSignalofRPC}:

\begin{equation}
    \tau = \frac{\epsilon}{\sigma} = \epsilon \rho
\end{equation}

Here $\epsilon, \sigma, \rho$ represent the dielectric constant, conductivity, and resistivity of the resistive plate, respectively. When the relative dielectric constant of the glass is 4.2 and the resistivity is $10^{10}$ $\Omega \cdot cm $, $\tau \sim$ 3.7 ms, which means that the avalanche charge accumulated in the resistive plate needs $\sim$ ms to be fully released. The lower the resistivity, the smaller the $\tau$, the faster the charge is released and the better the rate capability. Compared to traditional resistive plates $10^{12} \sim 10^{14}$ $\Omega \cdot cm$, the low resistive float glass($1.4\cdot 10^{10}$ $\Omega \cdot cm$) improves the rate capability by $2\sim4$ orders of amplitude. In the subsequent test, the low resistive float float glass was used uniformly.


\subsection{System time resolution}

\begin{figure}[htbp]
\centering
\subfloat[]{%
    \includegraphics[width=0.5\linewidth]{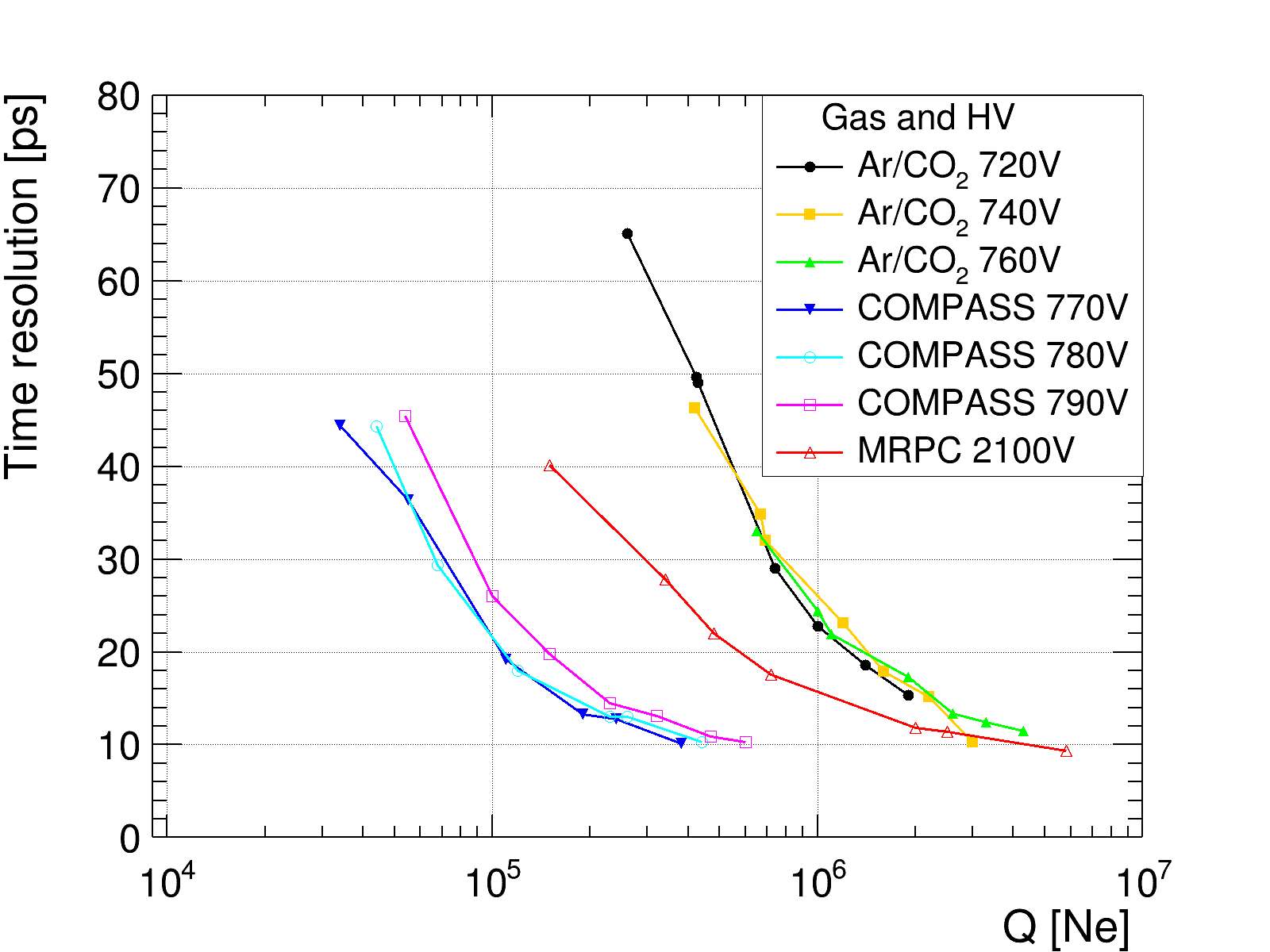}
    \label{ConstHV}
}
\subfloat[]{%
    \includegraphics[width=0.5\linewidth]{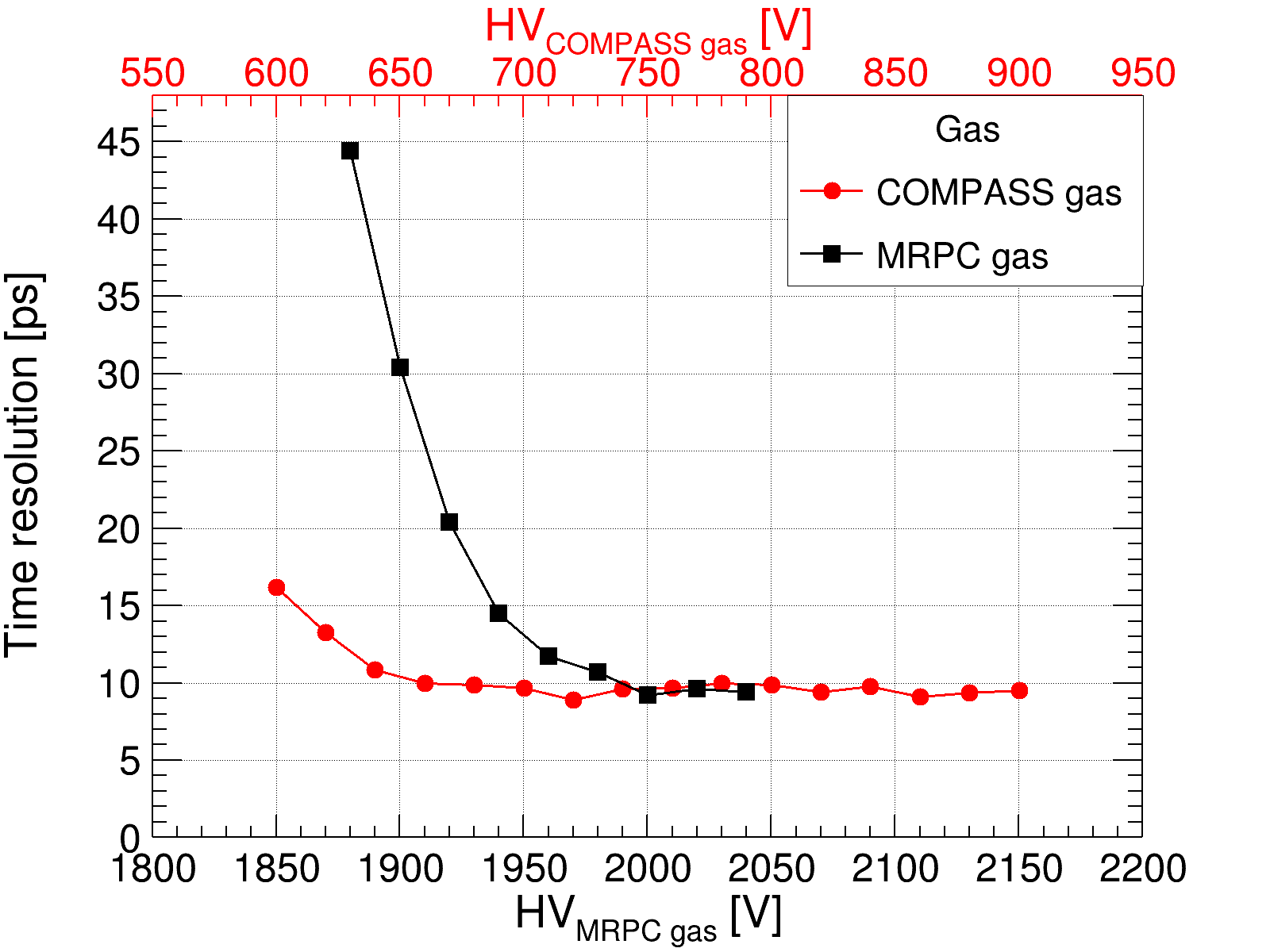}
    \label{ConstElaser}
}
\caption{Time resolution in different gases. (a) Keep the voltage constant and increase the incident laser intensity. (b) $E_{laser}$ = const, increase voltage.}
\label{TestSystemT}
\end{figure}

In the test, the PMT voltage was set at 280 V and the signal amplitude was maintained at 45 $\sim$ 60 mV by adjusting attenuator 2. In order to measure the time resolution of RPC, it is necessary to know the time resolution of the test system (mainly contributed by the PMT). Two methods are used to test the time resolution of the system: 1) under constant voltage, adjust the attenuator 1 to slowly increase the laser intensity and test the time resolution of the system. 2) Adjust the attenuator 1 so that the photocathode can produce enough photoelectrons ($\sim$ 100), then change the voltage and test the time resolution of the system. The test results are shown in Fig. \ref{TestSystemT}, in which we find that the time resolution of the test system is $\sim$ 10 ps regardless of increasing the laser intensity (increasing the initial number of photoelectrons) or increasing the applied electric field. Therefore, it is concluded that the time uncertainty of the test system (PMT+ amplifier + oscilloscope) under such conditions is $\sigma_{system}$ = 10 ps, which also means that when the initial number of photoelectrons is sufficient, the time resolution of the detector can be better than 10 ps.

\subsection{Test with single photoelectron}

The single photoelectron signal of the RPC was tested in different gases. Adjust attenuator 1 so that the frequency of the detector signal is much smaller than the frequency of the PMT signal ($\frac{f_{RPC}}{f_{PMT}} < 0.1$ ), at which point the RPC signal is treated as a single photoelectron signal. Oscilloscope was used to obtain waveform data (time, amplitude information) of RPC and PMT.

\begin{figure}[htbp]
    \centering
    \subfloat[]
    {
        \label{SEAr}
        \includegraphics[width=0.4\linewidth]{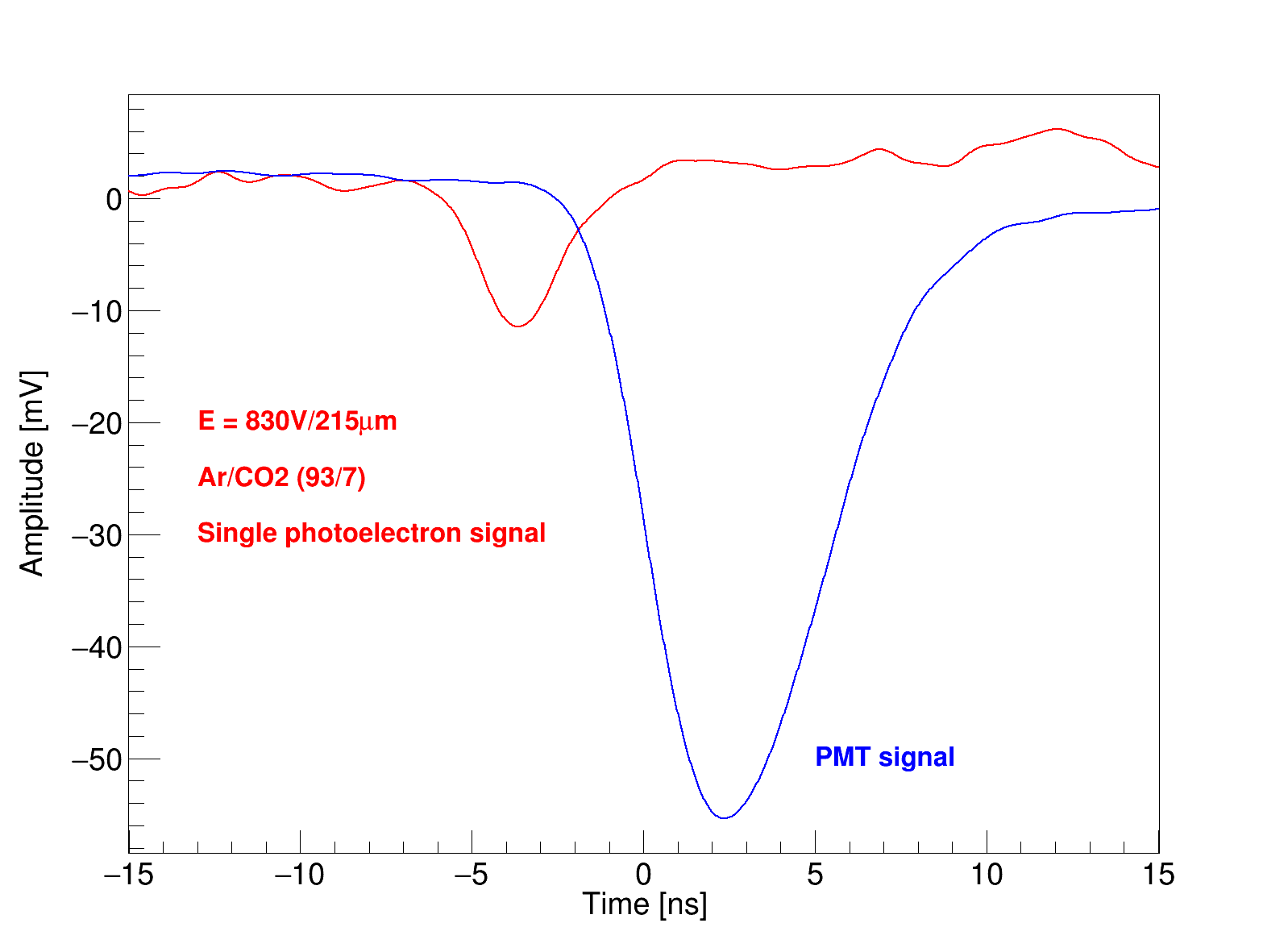}
    }
    \subfloat[]
    {
        \label{SENe}
        \includegraphics[width=0.4\linewidth]{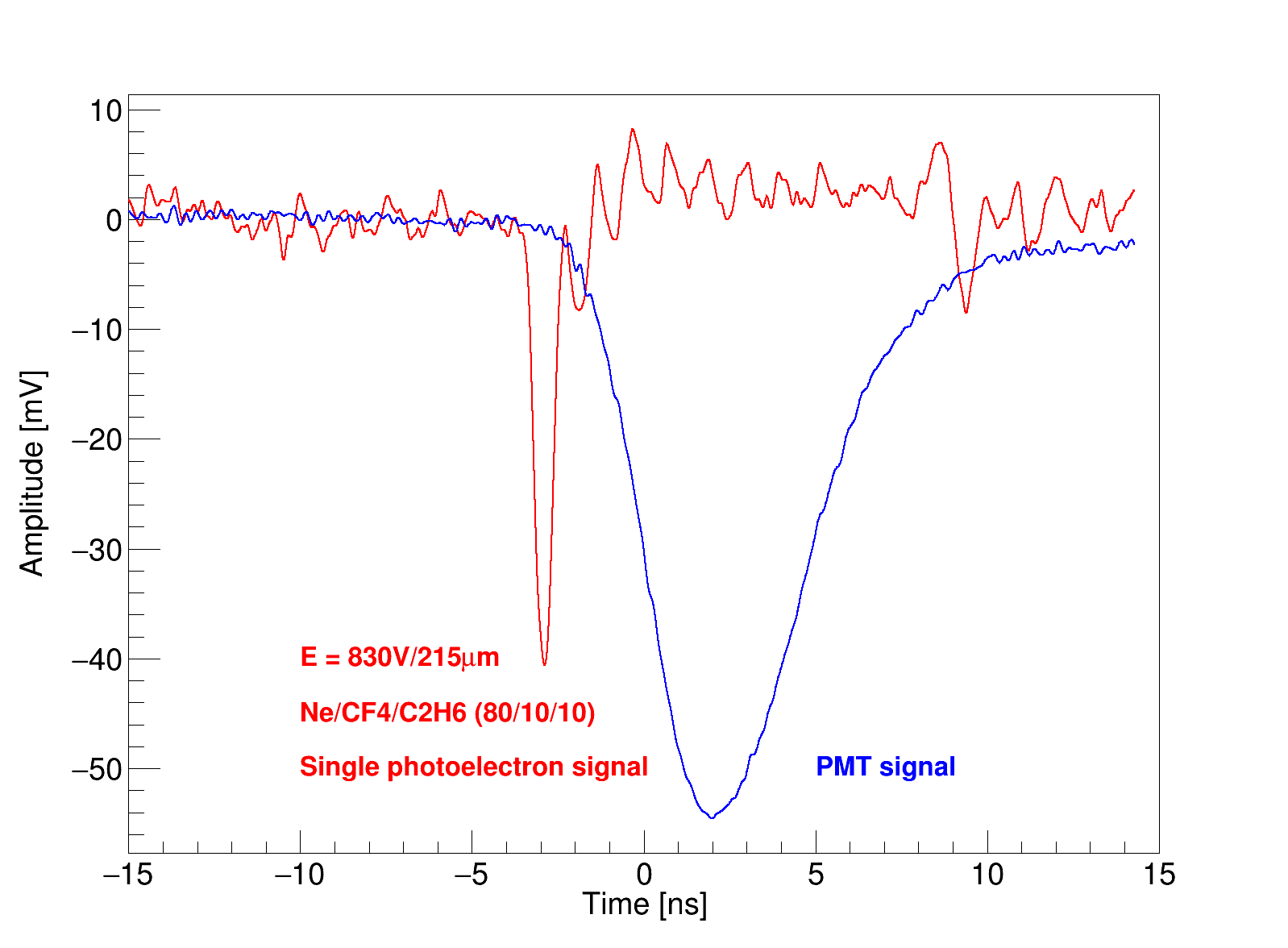}
    }
    \subfloat[]
    {
        \label{SER134a}
        \includegraphics[width=0.4\linewidth]{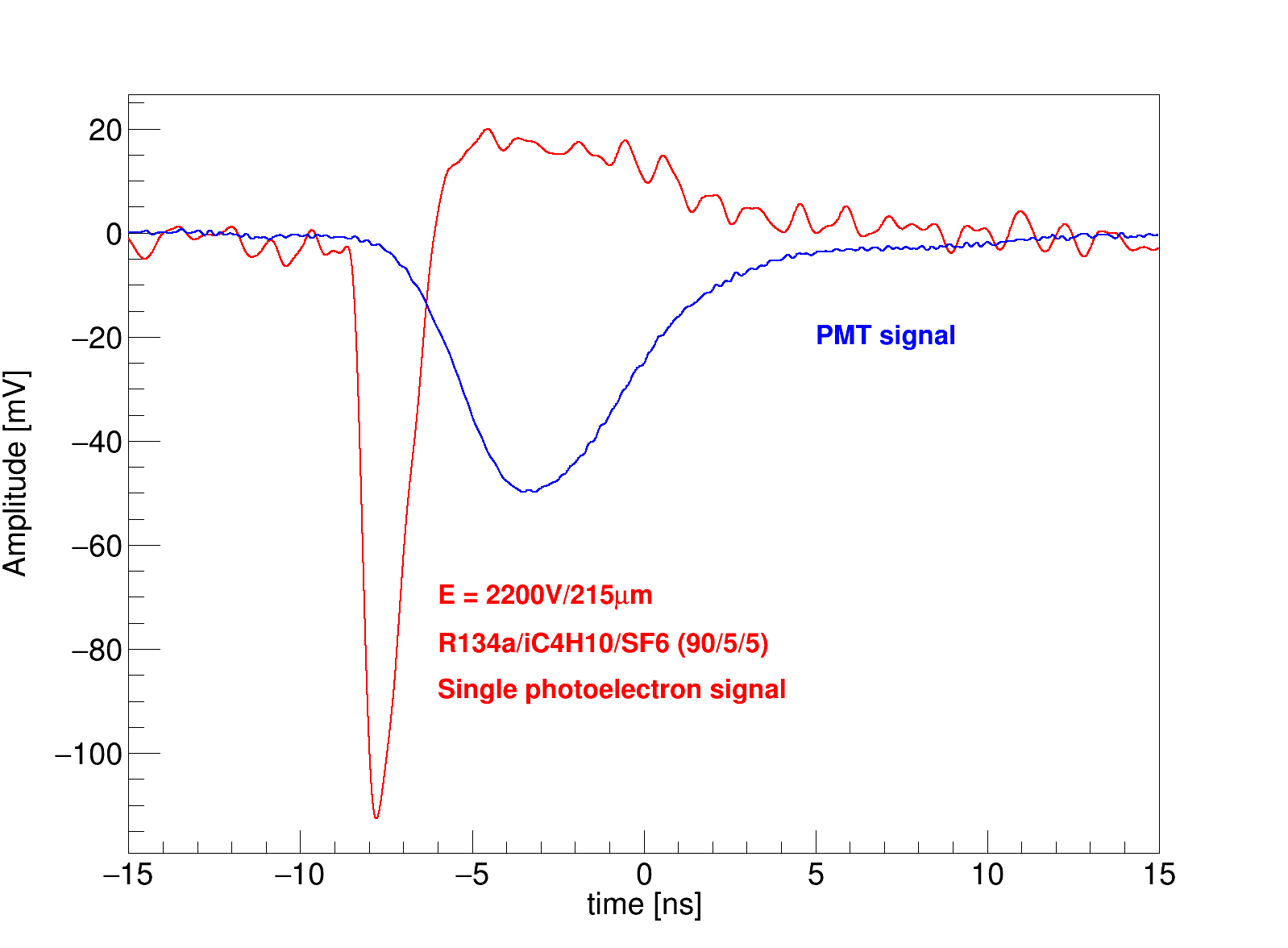}
    }
    \caption{ Signal collected by oscilloscope. Red curve represents single photoelectron signal of RPC (26 dB), while blue curve shows the synchronization signal generated by PMT. (a) $ArCO_{2}(93/7)$. (b) COMPASS gas. (c) MRPC gas.}
    \label{SEShape}
\end{figure}

 Fig. \ref{SEShape} shows the single photoelectron signal collected in different gases. In $Ar/CO_{2}$, the amplitude of the signal is small, $\sim$ 10 mV. Due to the small signal-to-noise ratio (S/N), the bandwidth limit of 2 GHz was adopted, and the maximum working voltage is $\sim$ 850 V. In COMPASS gas, the signal amplitude is larger, and the maximum working voltage is $\sim$ 840 V. In MRPC gas, the signal amplitude is the largest and the maximum working voltage is $\sim$2660 V. There is an overshoot caused by the amplifier along the trailing edge of the signal.

\begin{figure}
    \centering
    \subfloat[]{\label{SEA}\includegraphics[width=0.45\linewidth]{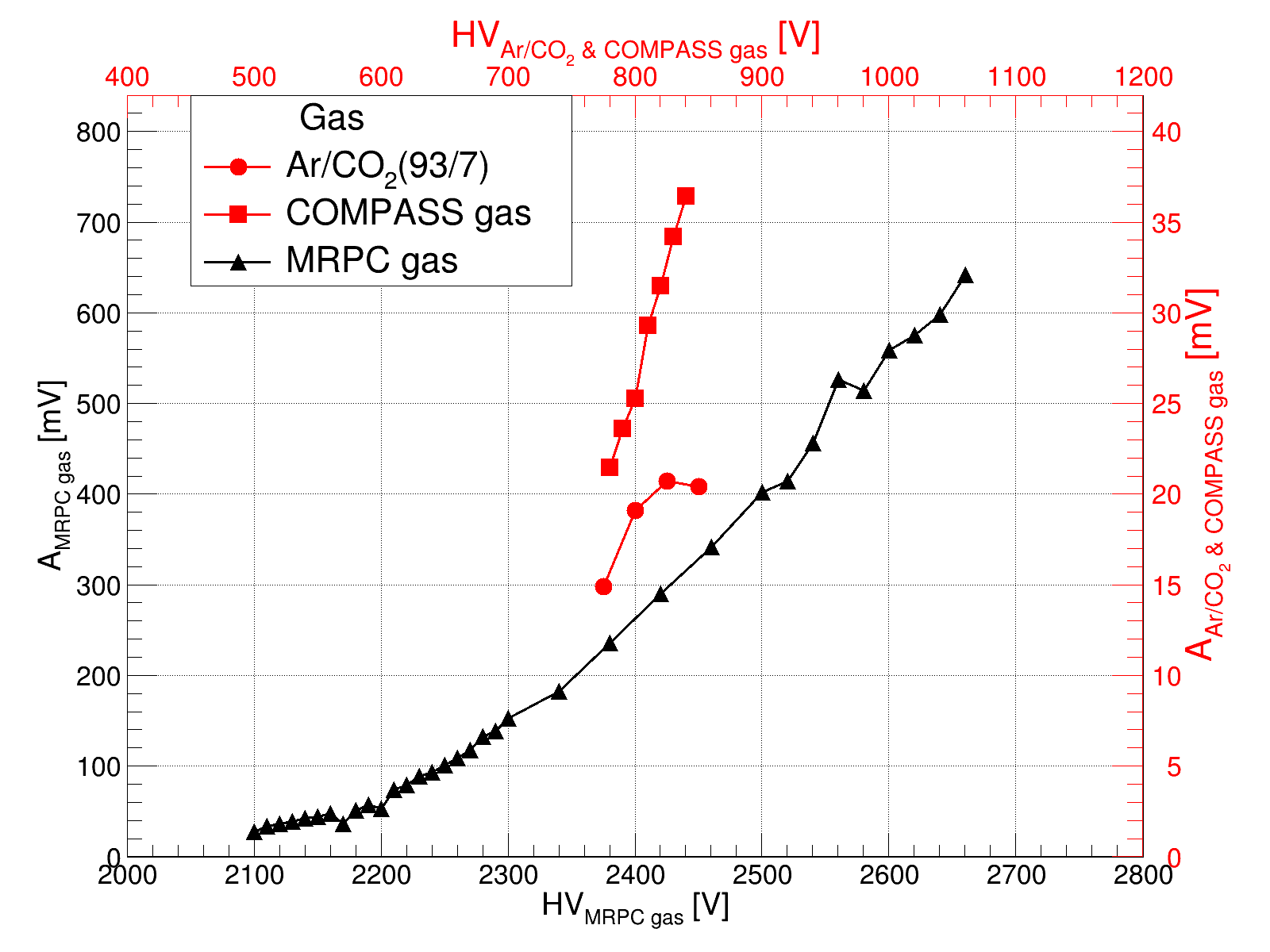}}
    \quad
    \subfloat[]{\label{SEFWHM}\includegraphics[width=0.45\linewidth]{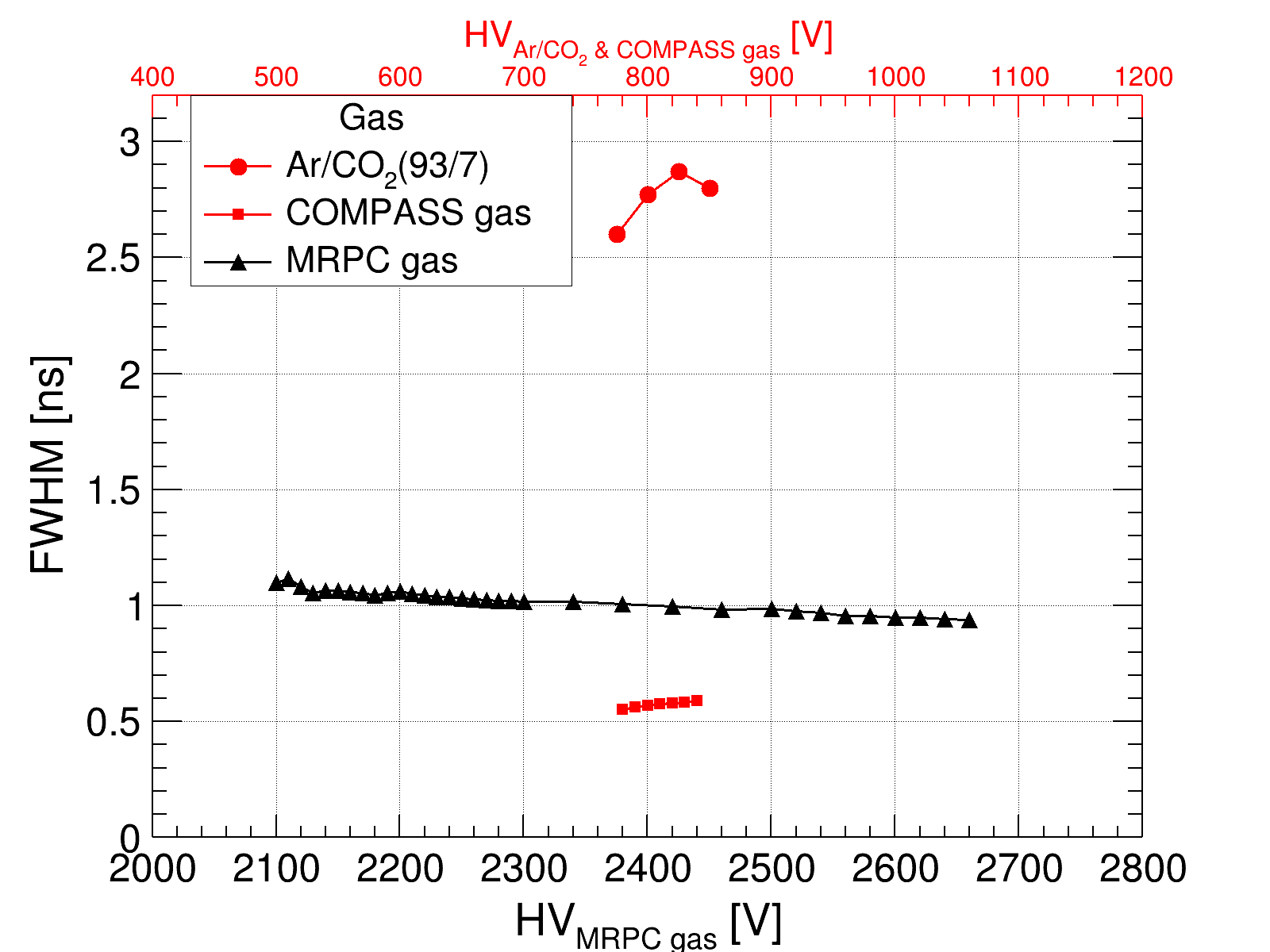}}
    \quad
    \subfloat[]{\label{SERT}\includegraphics[width=0.45\linewidth]{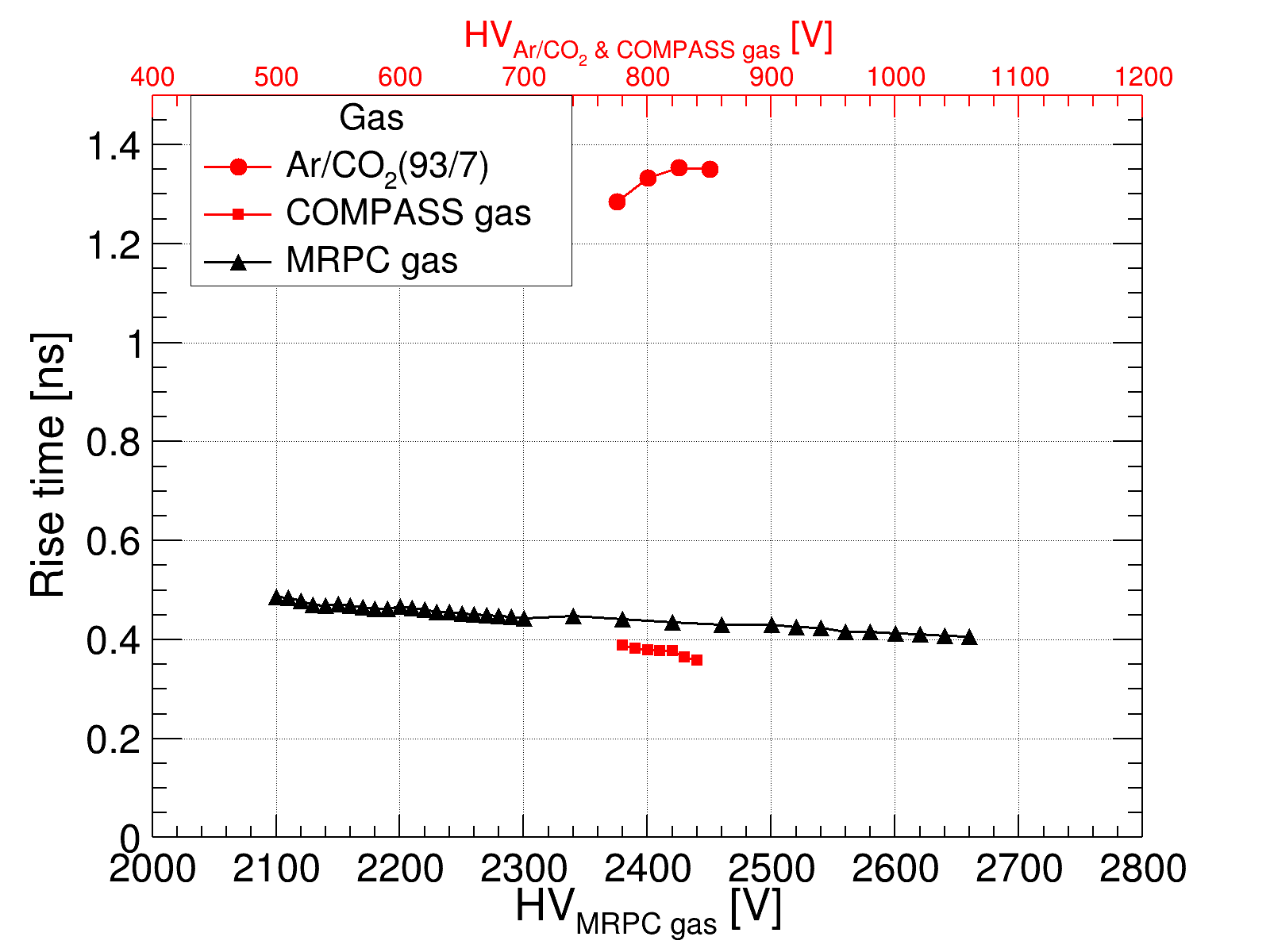}}
    \quad
    \subfloat[]{\label{SEASigmaRT}\includegraphics[width=0.45\linewidth]{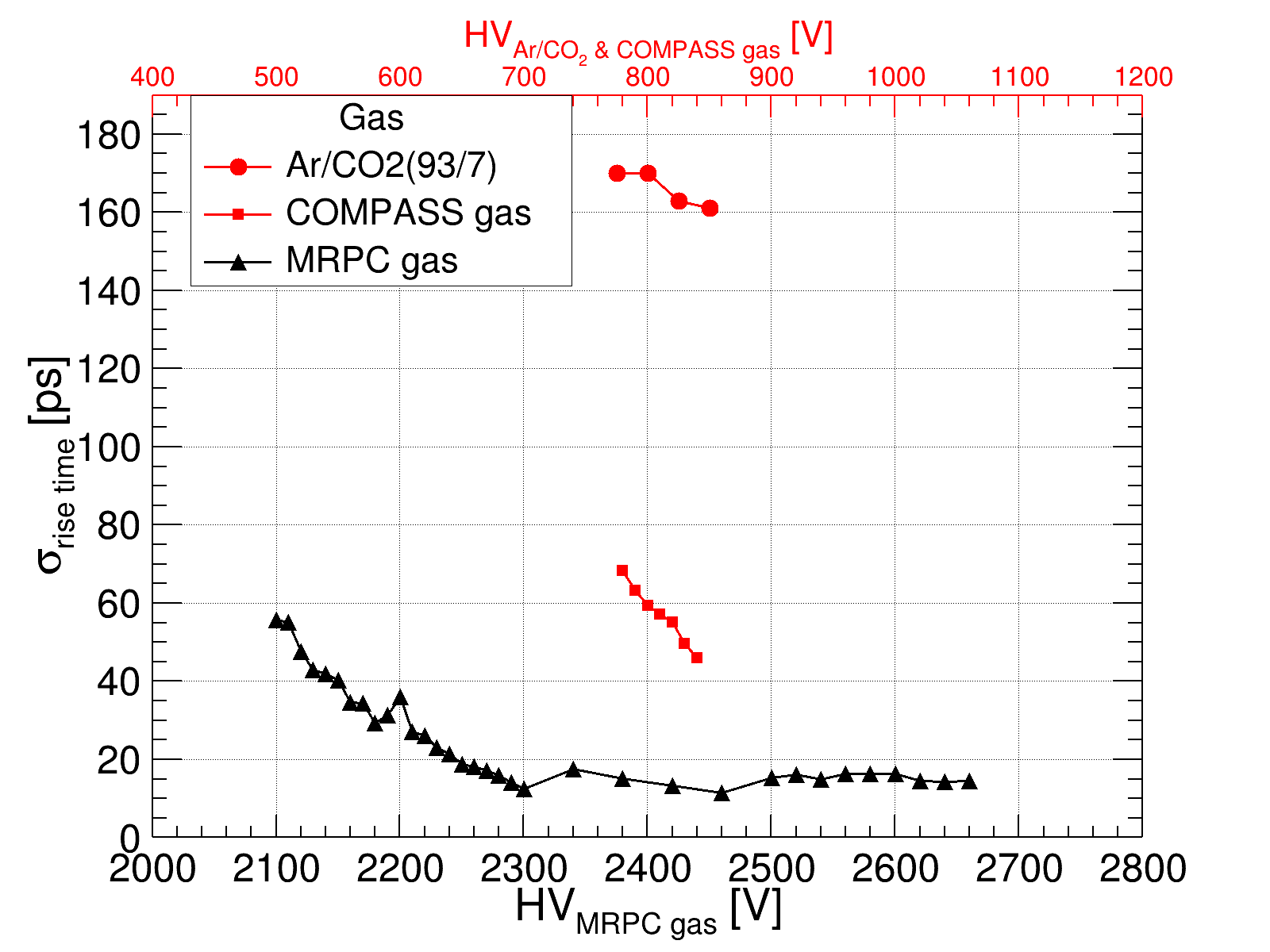}}
    \quad
    \caption{The characteristics of single photoelectron signals in different gases. (a) The amplitude (26 dB). (b) signal width (denoted by FWHM). (c) Rise time. (d) The spread of rise time $\sigma_{rise\ time}$ (the standard deviation of rise time).}
    \label{SignalFeature}
\end{figure}

The single photoelectron signals in $Ar/CO_{2}$ (775 V $\sim$ 850 V), COMPASS gas (780 V $\sim$ 840 V) and MRPC gas (2100 V $\sim$ 2660 V) were tested respectively. The amplitude, rise time ($|T_{0.1A_{max}} - T_{0.9A_{max}}|$), width of signal (denoted by FWHM) are analyzed, as shown in Fig. \ref{SignalFeature}. It can be seen that the signal amplitude in MRPC gas is the largest, up to several hundreds of mV. In the other two gases, the signal amplitude is only a few tens of mV, especially in $Ar/CO_{2}$, where the signal is the smallest and the S / N is the worst. The signal widths are about 2.8 ns, 0.6 ns and 1 ns, respectively. The mean value of rise time (corresponding to electron drift velocity) are about 1.3 ns, 0.38 ns and 0.45 ns, respectively. The standard deviation of the distribution of rise time $\sigma_{rise time}$ are $\sim$ 165 ps, 46 $\sim$ 68 ps, and 14 $\sim$ 56 ps, respectively, which means that the spread of the electron drift velocity is minimal in MRPC gas.

\subsection{The time resolution of single photoelectron}

\begin{figure}
    \centering
    \subfloat[]{\label{CFPMT}\includegraphics[width=0.5\linewidth]{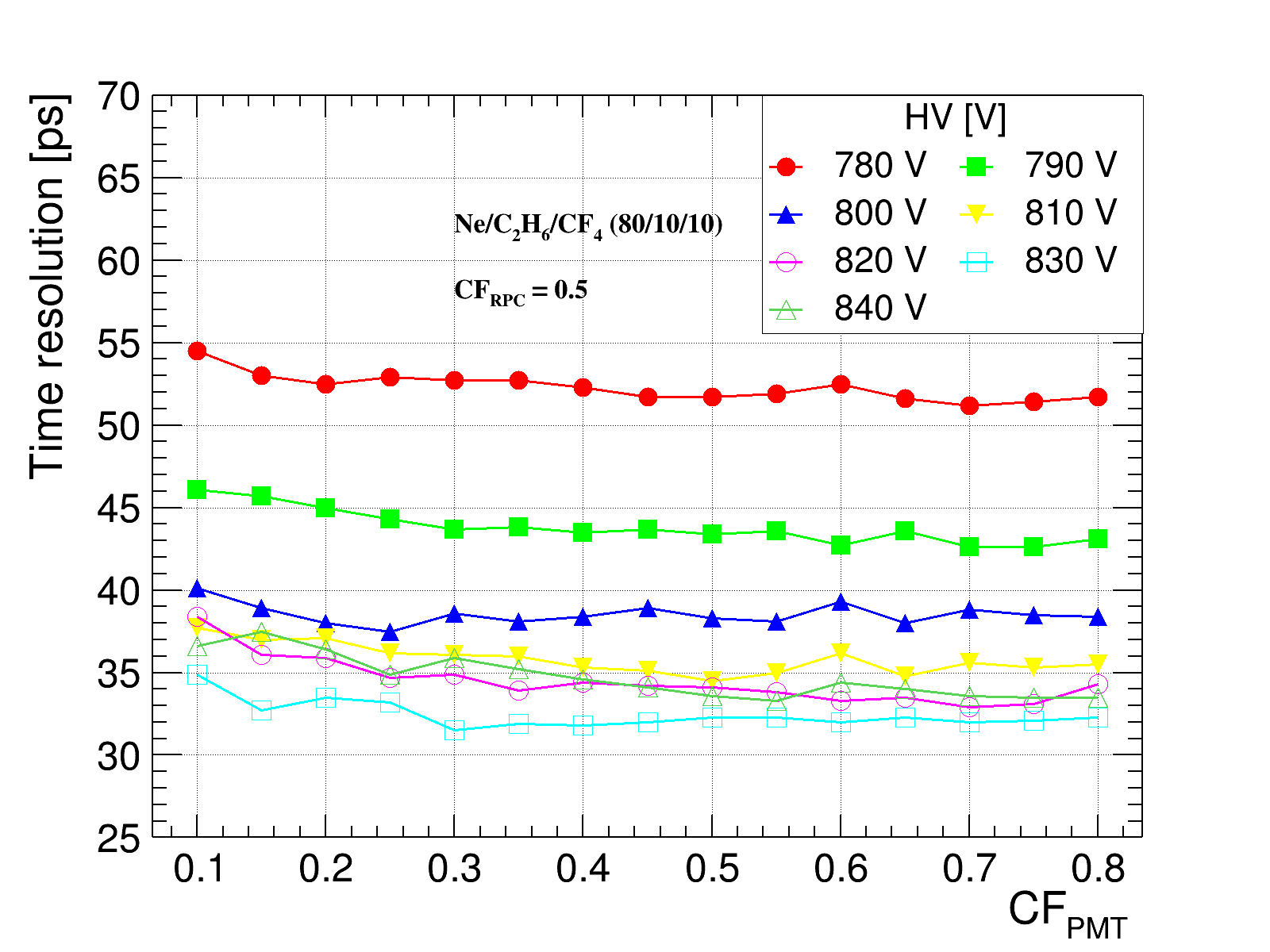}}
    \subfloat[]{\label{CFRPC}\includegraphics[width=0.5\linewidth]{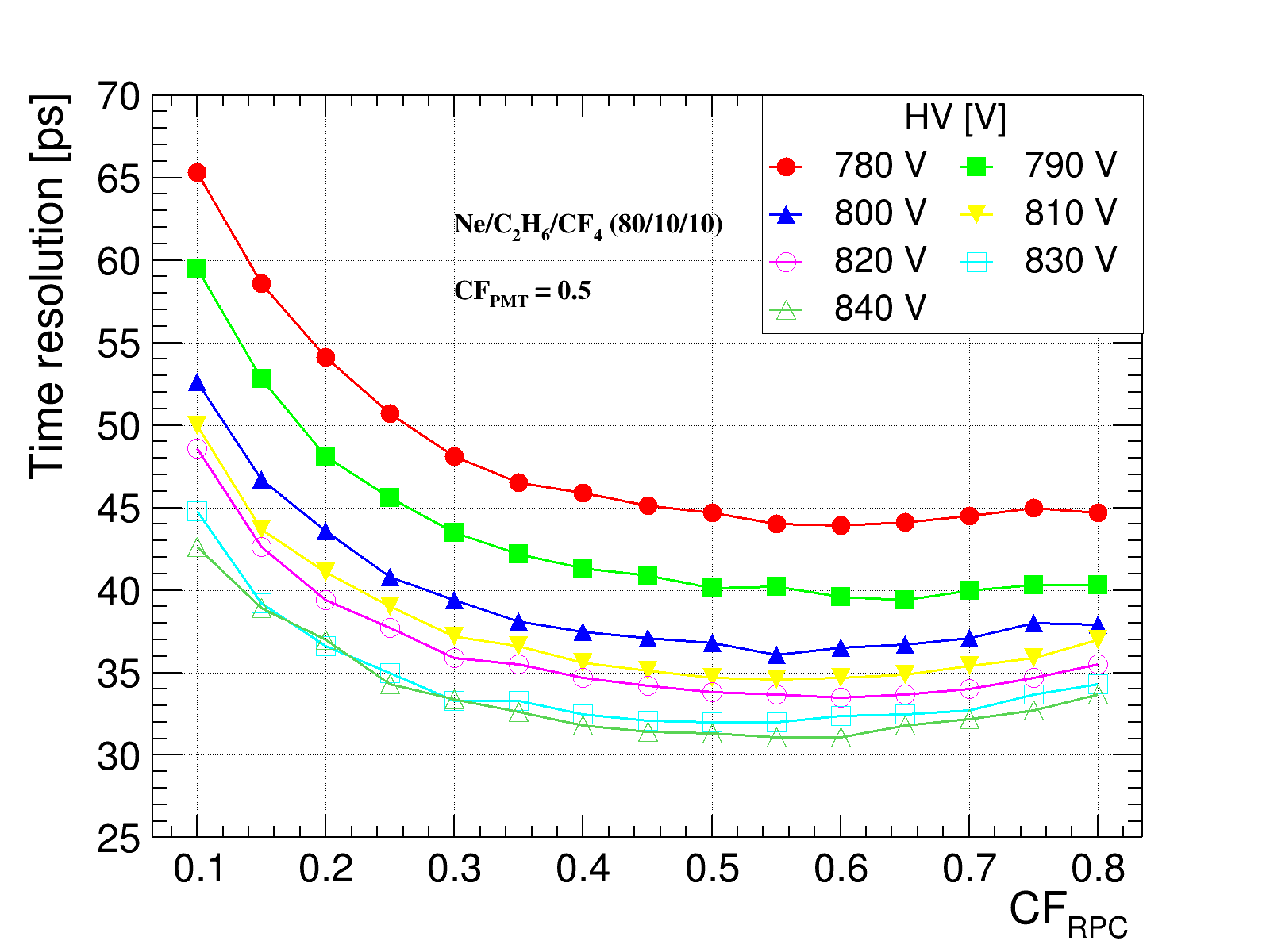}}
    
    \caption{Time resolution vs. CF value in COMPASS gas. (a) $CF_{RPC}$ = 0.5. (b) $CF_{PMT}$ = 0.5.}
    \label{CF}
\end{figure}

In a single photoelectron condition, the total time resolution $\sigma_{total}$ is calculated by constant-fraction (CF) discrimination(CFD), the time resolution of RPC is $\sigma_{RPC} = \sqrt{\sigma_{total}^{2} - \sigma_{system}^{2}}$. The optimal CF coefficients of the RPC and PMT of the two signals are determined by the following step: first, fix $CF_{RPC}$ to 0.5, change $CF_{PMT}$ from 0.1 to 0.9, and the value $CF_{PMT}$ is determined when the time resolution is optimal. Second, fixing the value of $CF_{PMT}$, change $CF_{RPC}$ from 0.1 to 0.9, and the value of $CF_{RPC}$ that makes the time resolution optimal is obtained. Take COMPASS gas as an example, the change of time resolution with CF coefficient is shown in Fig. \ref{CF}.

As can be seen from Fig.\ref{CFPMT}, for the PMT signal, the change in the CF value has little effect on time resolution, since the PMT signal is very large (signal amplitude $\sim$45 mV even without an amplifier), noise has little effect on it. The $CF_{PMT}$ was fixed at 0.5 in the subsequent data analysis. For RPC signals, as the CF value increases, the time resolution gradually becomes better until it reaches a flat area. This is because the amplitude of RPC signals is modest in the COMPASS gas, so the S/N is small. The noise has a significant influence on the timing. When $CF_{RPC} \sim$ 0.55, the time resolution is optimal.

\subsubsection{Performance in $Ar/CO_{2}$(93/7)}

In $Ar/CO_{2}$, the signal amplitude of the single photoelectron is limited. When the voltage is increased to 850 V, an obvious streamer signal was observed. At HV $\sim$ 840 V, the average amplitude is about 20 mV (Fig. \ref{SEA}), the S/N is small, and noise has an obvious influence on the shape of the signal. 



\begin{figure}
    \centering
    \subfloat[]{\label{6mV}\includegraphics[width=0.45\linewidth]{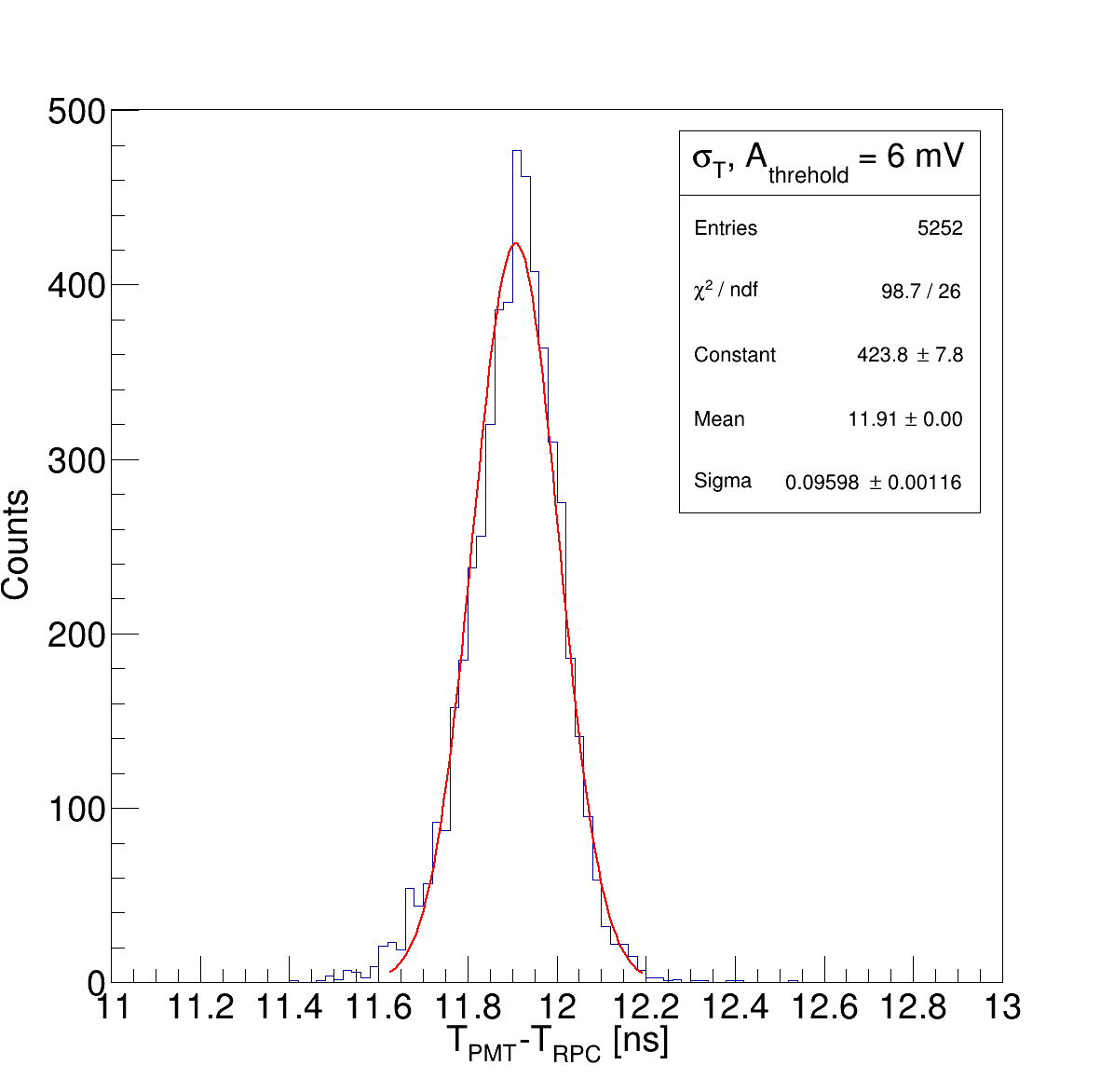}}
    \quad
    \subfloat[]{\label{10mV}\includegraphics[width=0.45\linewidth]{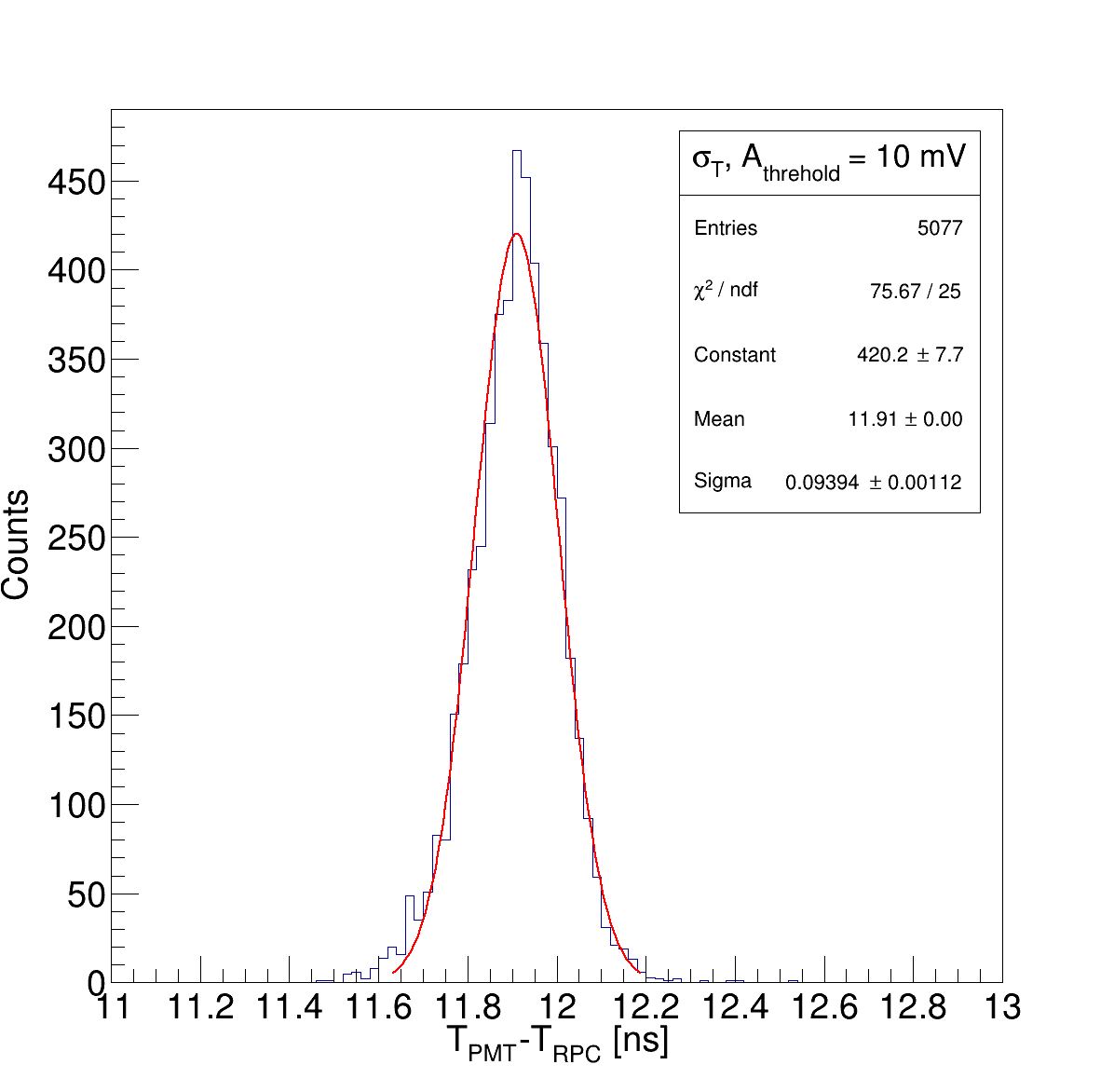}}
    \quad
    \subfloat[]{\label{14mV}\includegraphics[width=0.45\linewidth]{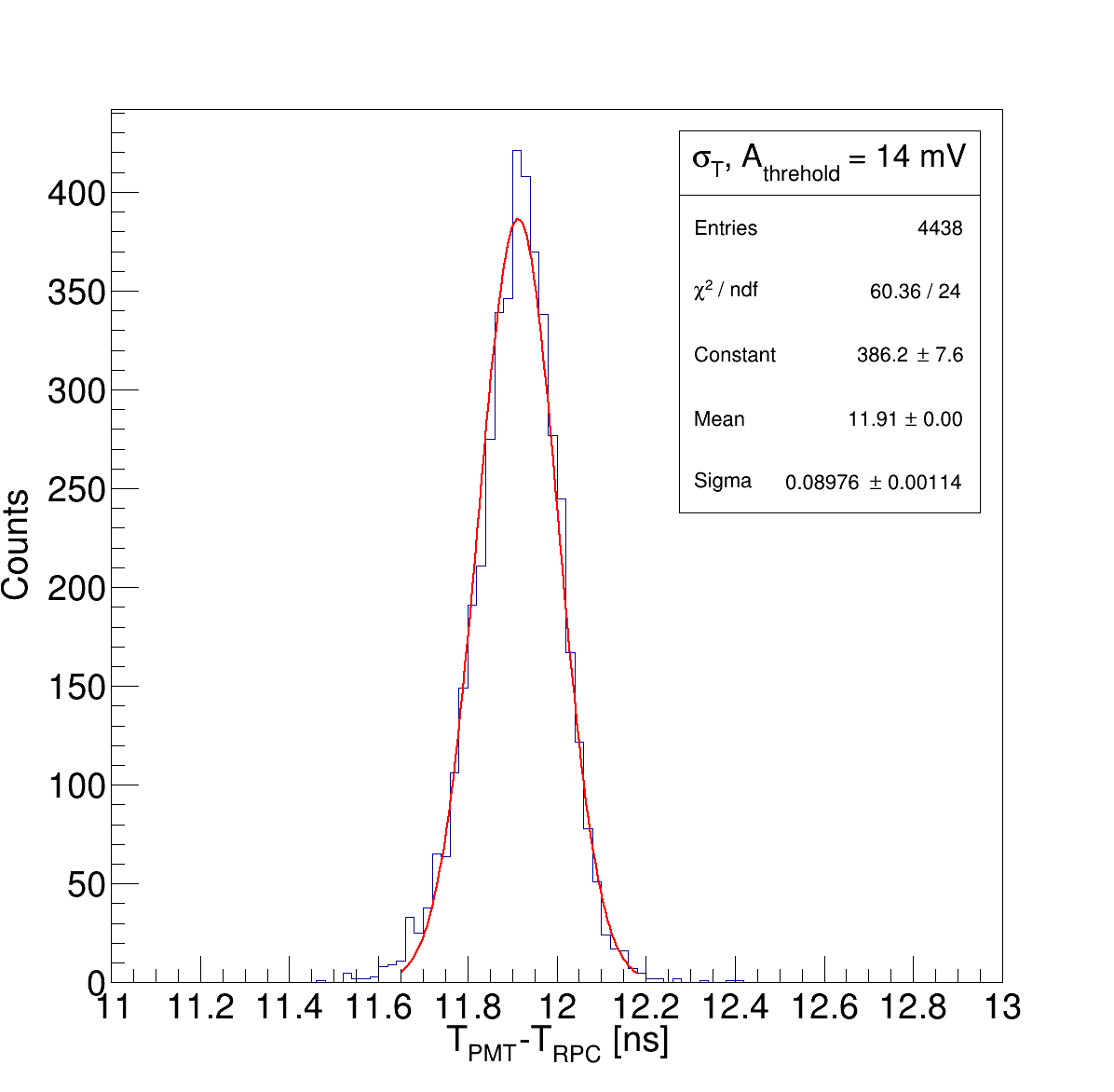}}
    \quad
    \subfloat[]{\label{18mV}\includegraphics[width=0.45\linewidth]{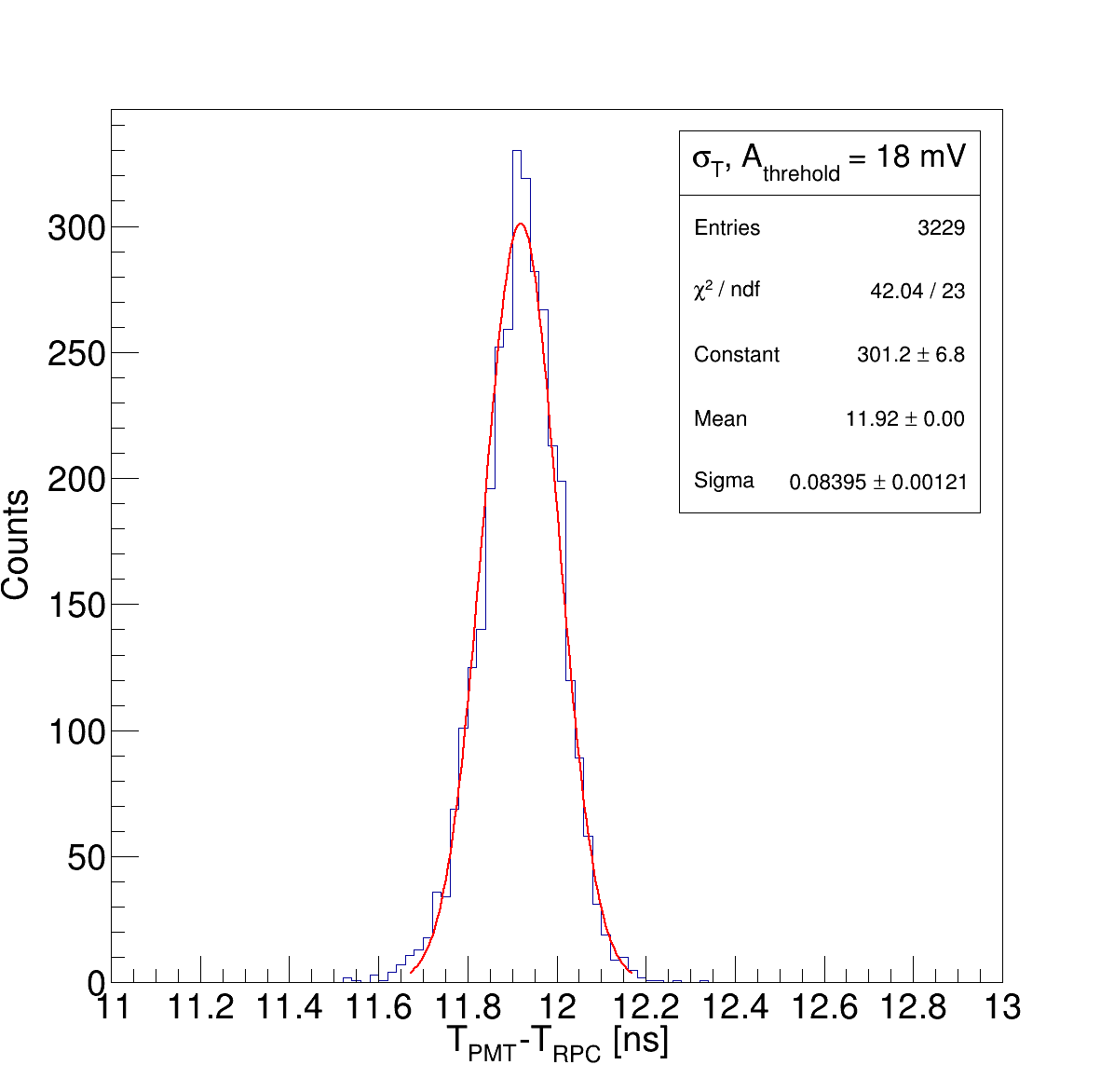}}
    
    \caption{The time resolution when the threshold is 6 mV(a), 10 mV(b), 14 mV(c), 18 mV(d) at HV = 825V, $CF_{RPC}$ = 0.55 in $Ar/CO_{2}$(93/7).}
    \label{SESigmaT825}
\end{figure}

Since the S/N is not good, it is difficult to perform timing analysis for all signals, the time resolution under different threshold cuts ($>$ 6, 10, 14, 18 mV) was calculated. Fig.\ref{SESigmaT825} shows the time resolution (95.6 ps, 93.9 ps, 89.8 ps, 84.0 ps) under different threshold cuts. It is clear that as the threshold increases, the time resolution continues to improve (while of course the efficiency decreases), this is consistent with the expectation that the S/N has a significant influence on the time resolution.

The same data processing was done for the results at other voltages. Take 6 mV as the threshold, the results are shown in Table \ref{TabAr}.

\begin{table}[]
    \centering
    \caption{Time resolution and Q in $Ar/CO_{2}$(93/7)}
    \label{TabAr}
    \begin{tabular}{|c|c|c|}
        \hline
        \multicolumn{3}{|c|}{$CF_{RPC} = 0.55, A_{threhold} = 6 mV$} \\
        \hline
        HV[V] & Q [Ne] & Time resolution [ps] \\
        \hline
        775 & $2.3\cdot 10^{5}$ & 121.9 \\
        \hline
        800 & $4.4\cdot 10^{5}$ & 100 \\
        \hline
        825 & $5.5\cdot 10^{5}$ & 95.9 \\
        \hline
        850 & $4.8\cdot 10^{5}$ & 100.9 \\
        \hline
    \end{tabular}
\end{table}

In summary, this detector can working in $Ar/CO_{2}$(93/7) and get the best single photoelectron time resolution at 825 V, with a time resolution of $\sim$ 95.9 ps at a threshold of 6 mV. Improving the S/N (increasing the threshold or decreasing the noise) can improves the time resolution. Since the rise time ($\sim$ 1.35 ns) and signal width ($\sim$ 2.38 ns) are rather large (Fig. \ref{SignalFeature}), the signal amplitude is small ($<$ 20 mV), $Ar/CO_{2}$(93/7) is not a good working gas for this kind of photodetector.

\subsubsection{Performance in COMPASS gas - $Ne/CF_{4}/C_{2}H_{6} (80/10/10)$}

\begin{figure}
    \centering
    \subfloat[]{\label{QNe}\includegraphics[width=0.5\linewidth]{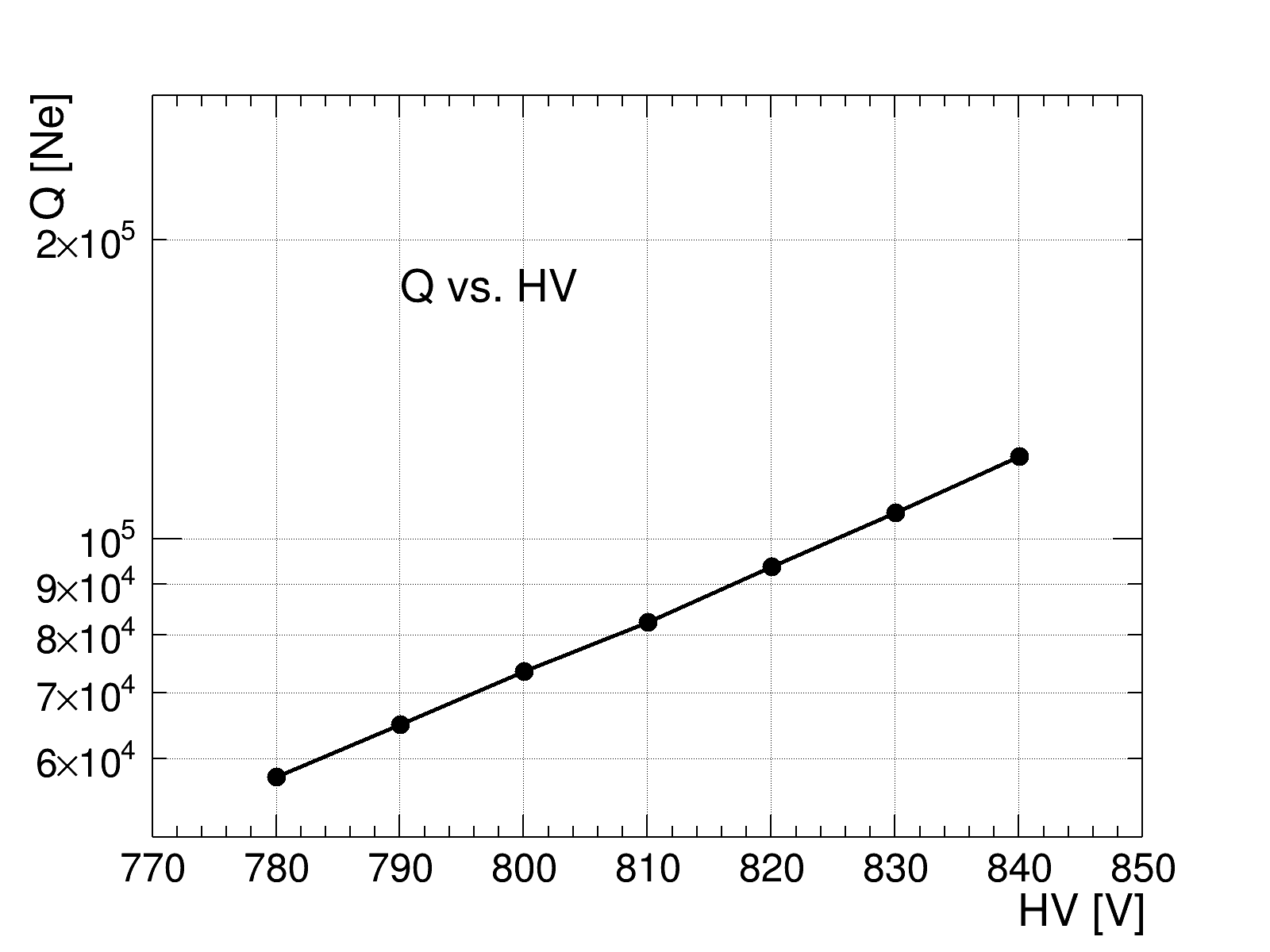}}
    \subfloat[]{\label{TNe}\includegraphics[width=0.5\linewidth]{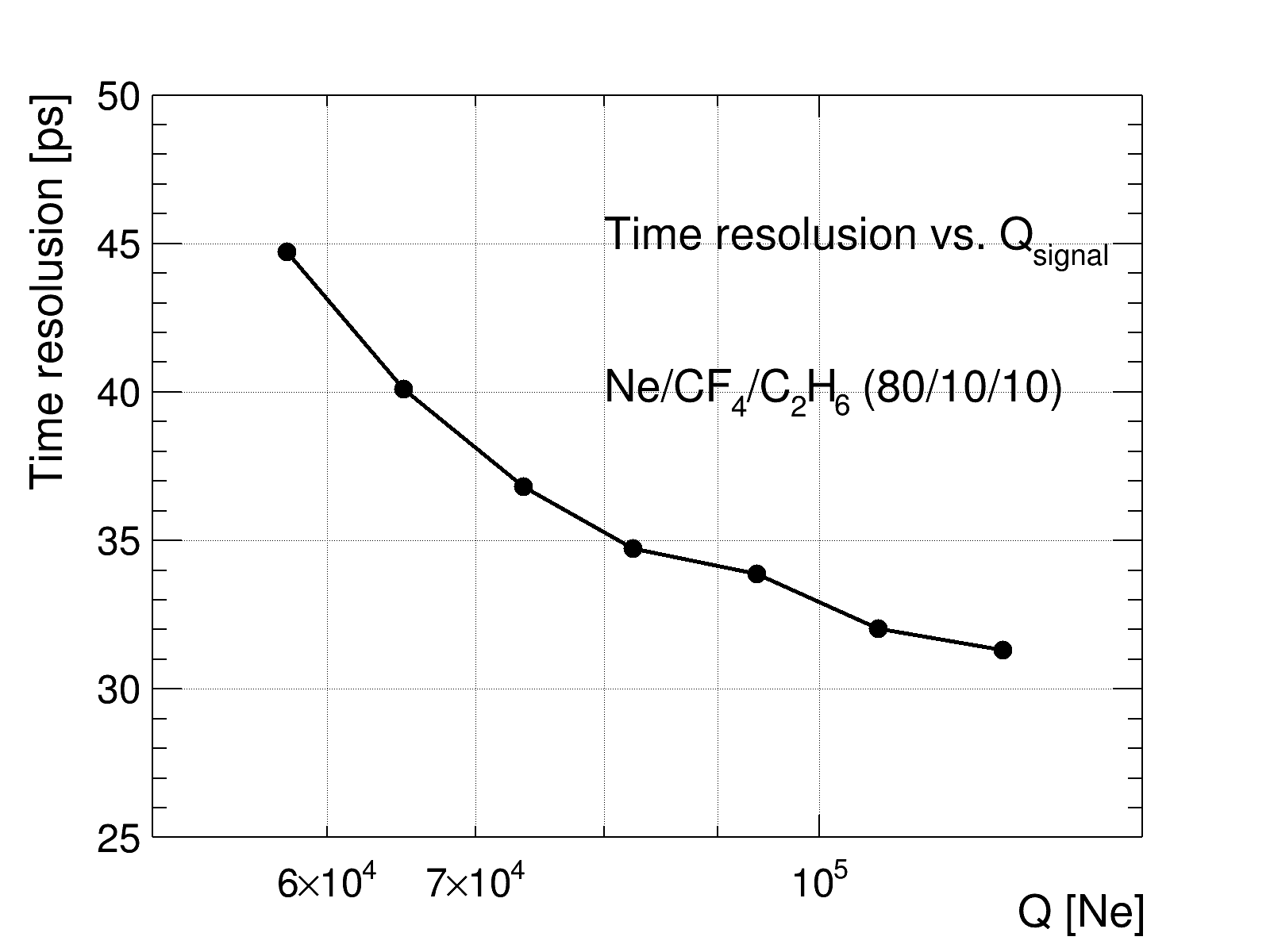}}
    
    \caption{Signal charge (a) and time resolution(b) in COMPASS gas.}
    \label{SigmaTNe}
\end{figure}

The signal quality in the COMPASS gas is significantly better than $Ar/CO_{2}$ (in terms of the S/N ratio, signal amplitude, rise time, and signal width, see Fig. \ref{SENe}, Fig. \ref{SignalFeature}). The signals obtained at different voltages are analyzed with the optimal $CF_{RPC}$ = 0.55. The signal charge (Q) and time resolution of single photoelectron are shown in Fig. \ref{SigmaTNe}. With an increase in voltage, the gain becomes larger and the resolution of time becomes better. At 840 V, the best time resolution is about 31.3 ps with a gain of $1.2\cdot 10^5$. When the voltage continues to increase to 850V, a clear streamer was seen.

In COMPASS gas, the photoelectric RPC can achieve a good single photoelectron time resolution ($\sim$ 31.3 ps) at a gas gain of $1.2\cdot 10^{5}$ at lower voltages (840 V).

\subsubsection{Performance in MRPC gas - $R134a/iC_{4}H_{10}/SF_{6}$}

Upon replacement of the working gas with a typical MRPC gas, $R134a/iC_{4}H_{10}/SF_{6} (90/5/5)$, a clear single photoelectron signal can be observed at 2100 V $\sim$ 2700 V (Fig. \ref{SER134a}). The amplitude of the signal is large, up to several hundreds mV (Fig. \ref{SEA}). There is no visible streamer until HV = 2750V.

\begin{figure}
    \centering
    \subfloat[]{\label{oneE}\includegraphics[width=0.5\linewidth]{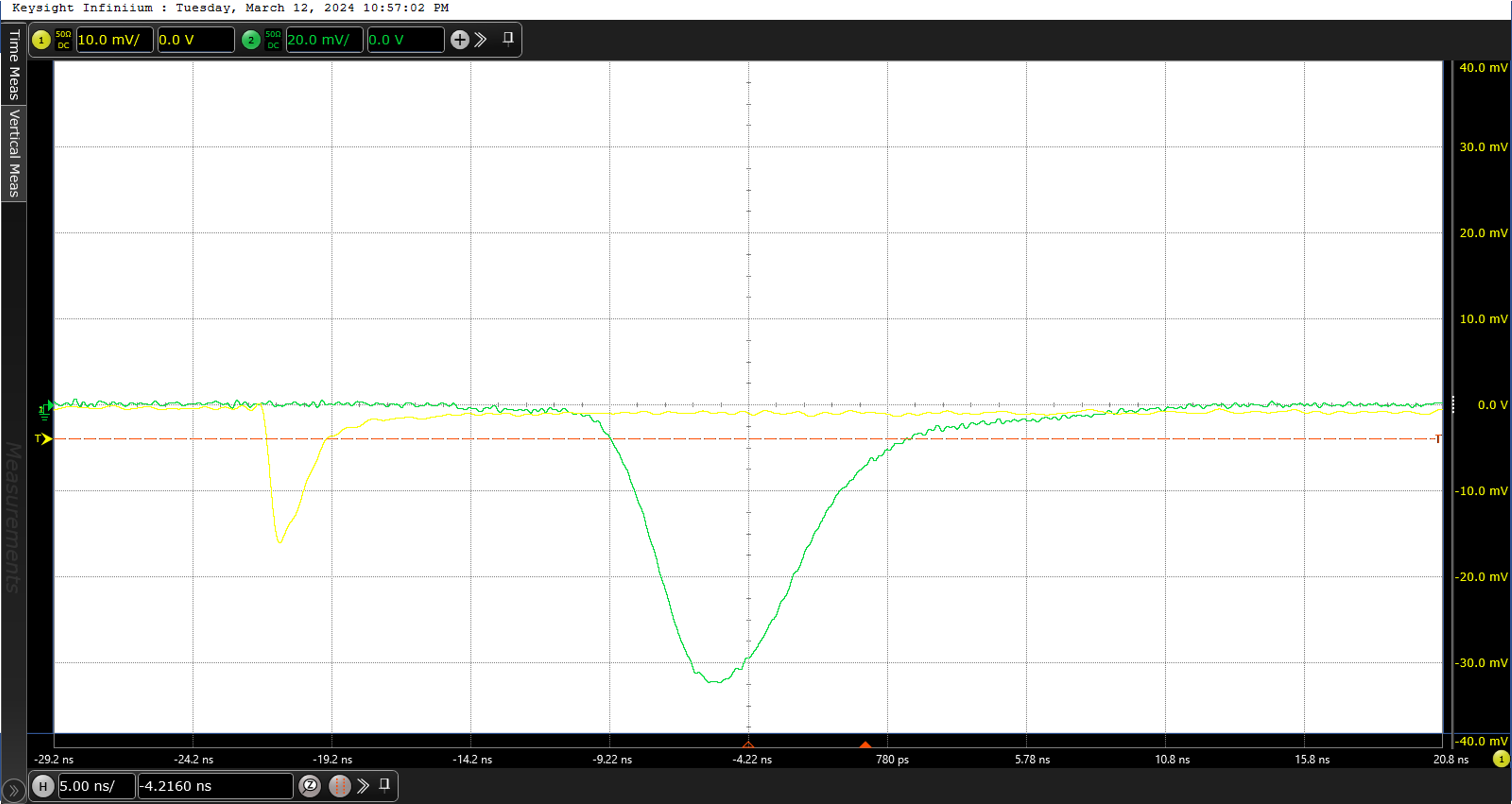}}
    \subfloat[]{\label{twoE}\includegraphics[width=0.5\linewidth]{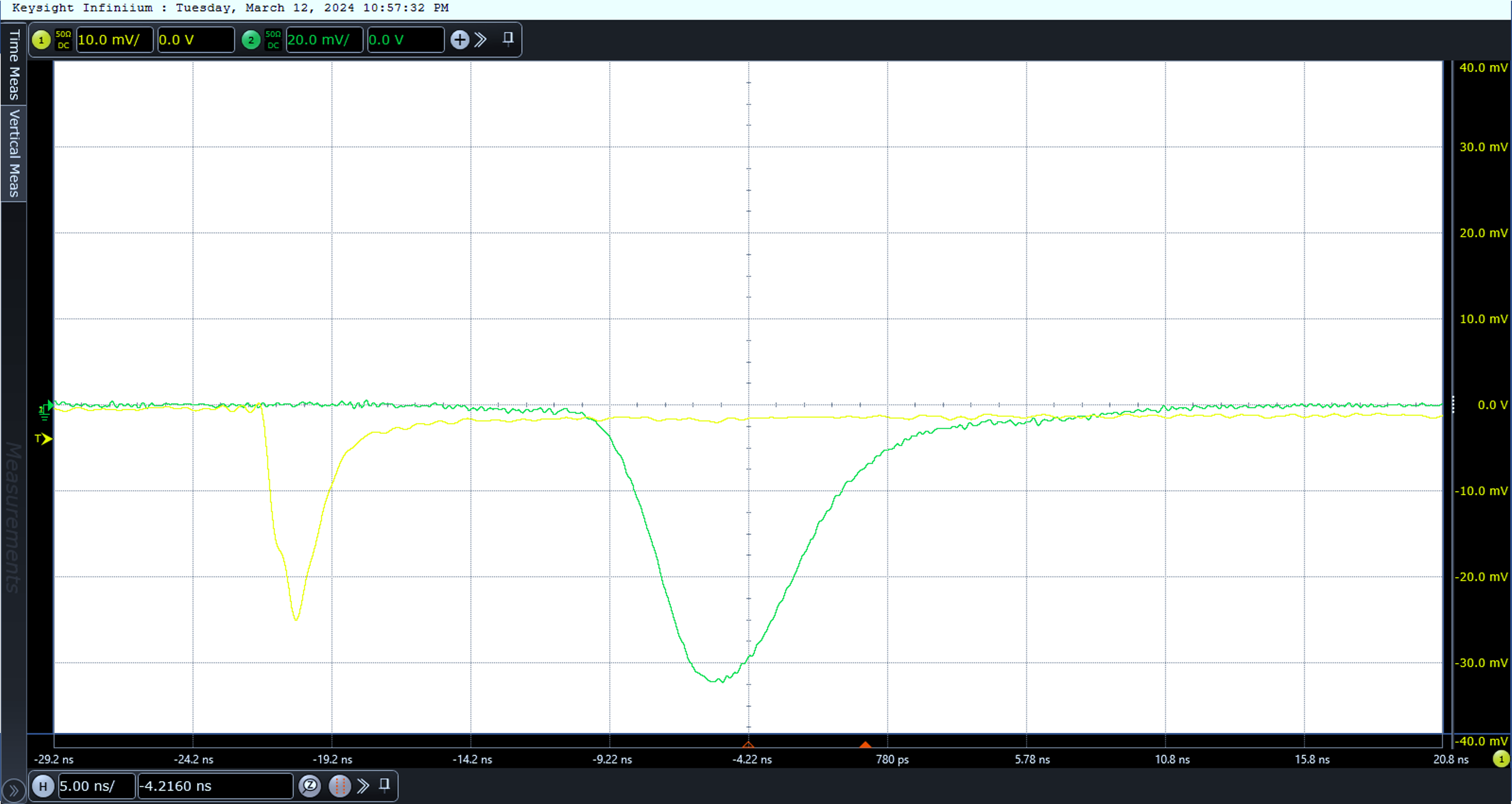}}
    
    \caption{Single photoelectron signal collected by oscilloscope at HV = 2500 V, without amplifier. (a) Single photoelectron signal. (b) The signal with secondary electrons generated by photon feedback.}
    \label{photo_SEsigna2500}
\end{figure}

When the voltage is high, the effect generated by photon feedback can be observed, as shown in Fig. \ref{photo_SEsigna2500}. The yellow curve indicates the RPC signal (without an amplifier) with an amplitude of 20 mV, the green curve is the PMT signal. Fig. \ref{oneE} is the signal for gas avalanche produced by one primary photoelectron, the leading-edge of the signal shape is smooth. An obvious turning point can be seen in Fig. \ref{twoE}, indicating a new signal is superimposed on the signal generated by the primary photoelectron, which is generated by photon feedback. When the gas gain is high, ultraviolet (UV) photons produced in the avalanche may escape and excite the photo-cathode or gas molecular far outside the initial avalanche to produce a new "primary" electron. A new avalanche may be triggered and the new signal will be superimposed on the original one.

\begin{figure}
    \centering
    \subfloat[]{\label{QvsRT2500}\includegraphics[width=0.5\linewidth]{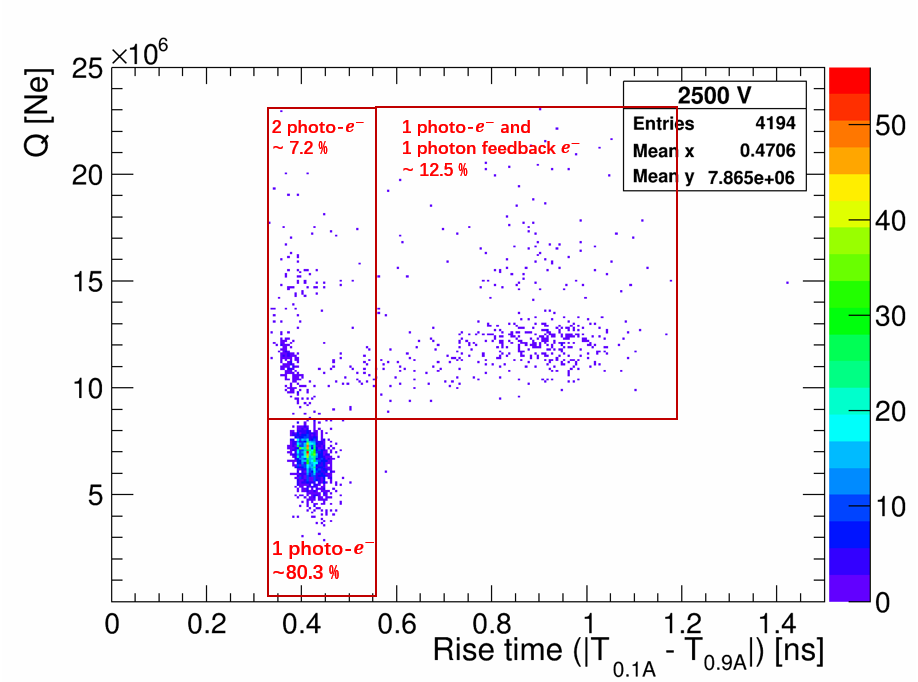}}
    \subfloat[]{\label{Q2500}\includegraphics[width=0.5\linewidth]{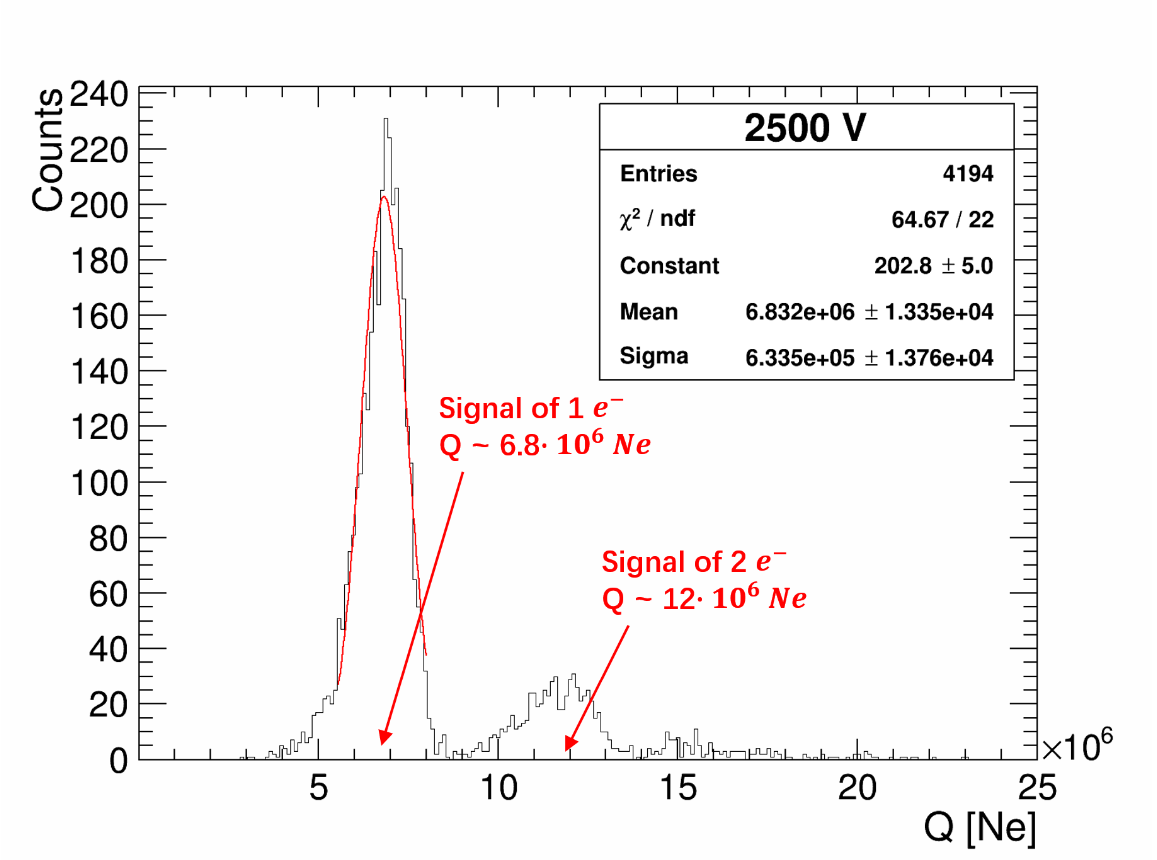}}
    \caption{Signals collected by oscilloscope (26 dB) at HV = 2500 V in MRPC gas ($R134a/iC_{4}H_{10}/SF_{6}(90/5/5)$). (a) The distribution of charge (Q) and rise time of signal. (b) The distribution of Q.}
    \label{SEsigna2500}
\end{figure}

The rise time and charge of the signals were analyzed in more detail. The signals collected at HV = 2500 V were taken as an example, as shown in Fig. \ref{SEsigna2500}, it can be clearly seen that the signals of single primary photoelectron with no photon feedback is concentrated in the vicinity of rise time $\sim$ 0.41 ns and Q $\sim 6.8\cdot10^{6}$ Ne, the proportion is 3368/4188 $\sim$ 0.80. The signals with 2 primary photoelectrons and the signals with photon feedback can be clearly distinguished in Fig. \ref{QvsRT2500}. Photon feedback makes the rise time larger, which worsens the timing performance.

\begin{figure}
    \centering
    \subfloat[]{\label{QR134a}\includegraphics[width=0.5\linewidth]{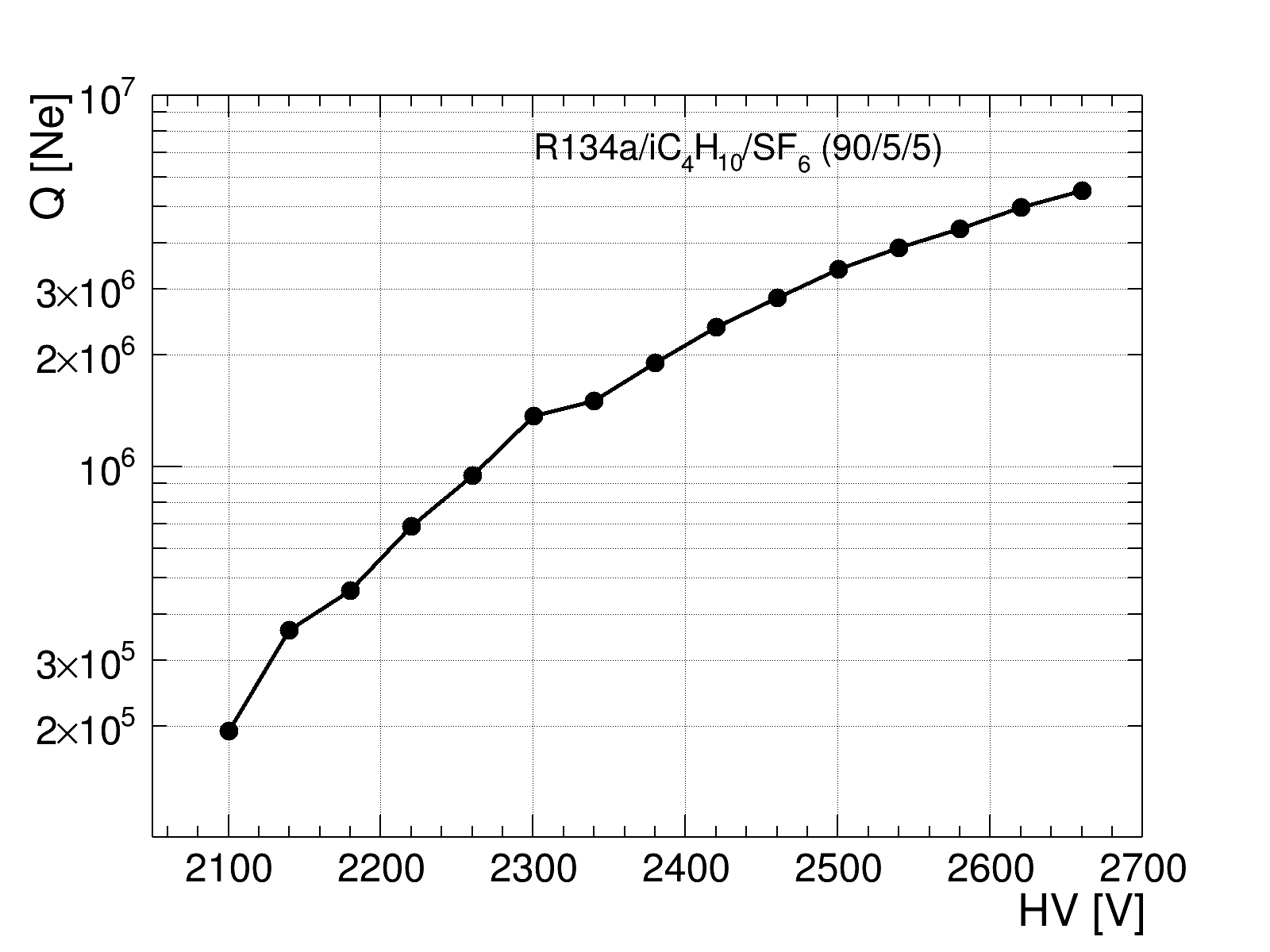}}
    \subfloat[]{\label{TR134a}\includegraphics[width=0.5\linewidth]{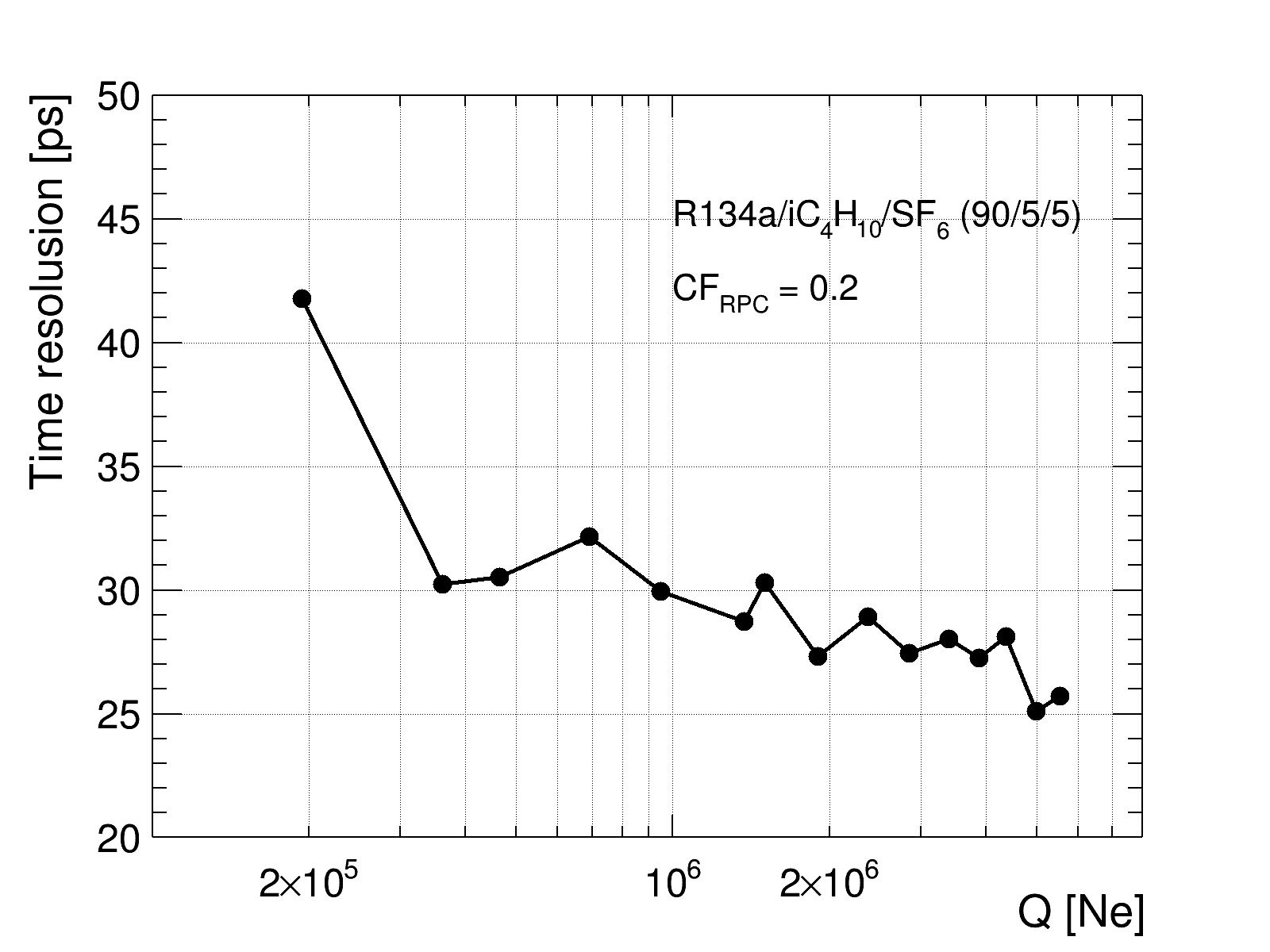}}
    
    \caption{Signal charge (a) and time resolution (b) in MRPC gas.}
    \label{SigmaTR134a}
\end{figure}

Only signals without photon feedback(Fig. \ref{oneE}) are counted, the charge and time resolution of a single photoelectron in MRPC gas are shown in Fig. \ref{SigmaTR134a}, with $CF_{RPC}$ = 0.2 for all voltages. It can be seen that the signal charge in MRPC gas is very large, up to $5\cdot 10^{6}$ Ne. Due to the space charge effect\cite{spaceCharge}, the charge is not exponentially dependent on the voltage. The time resolution of a single photoelectron can reach $\sim$ 25 ps at a gain of $5\cdot 10^{6}$.

\begin{figure}
    \centering
    \includegraphics[width=1\linewidth]{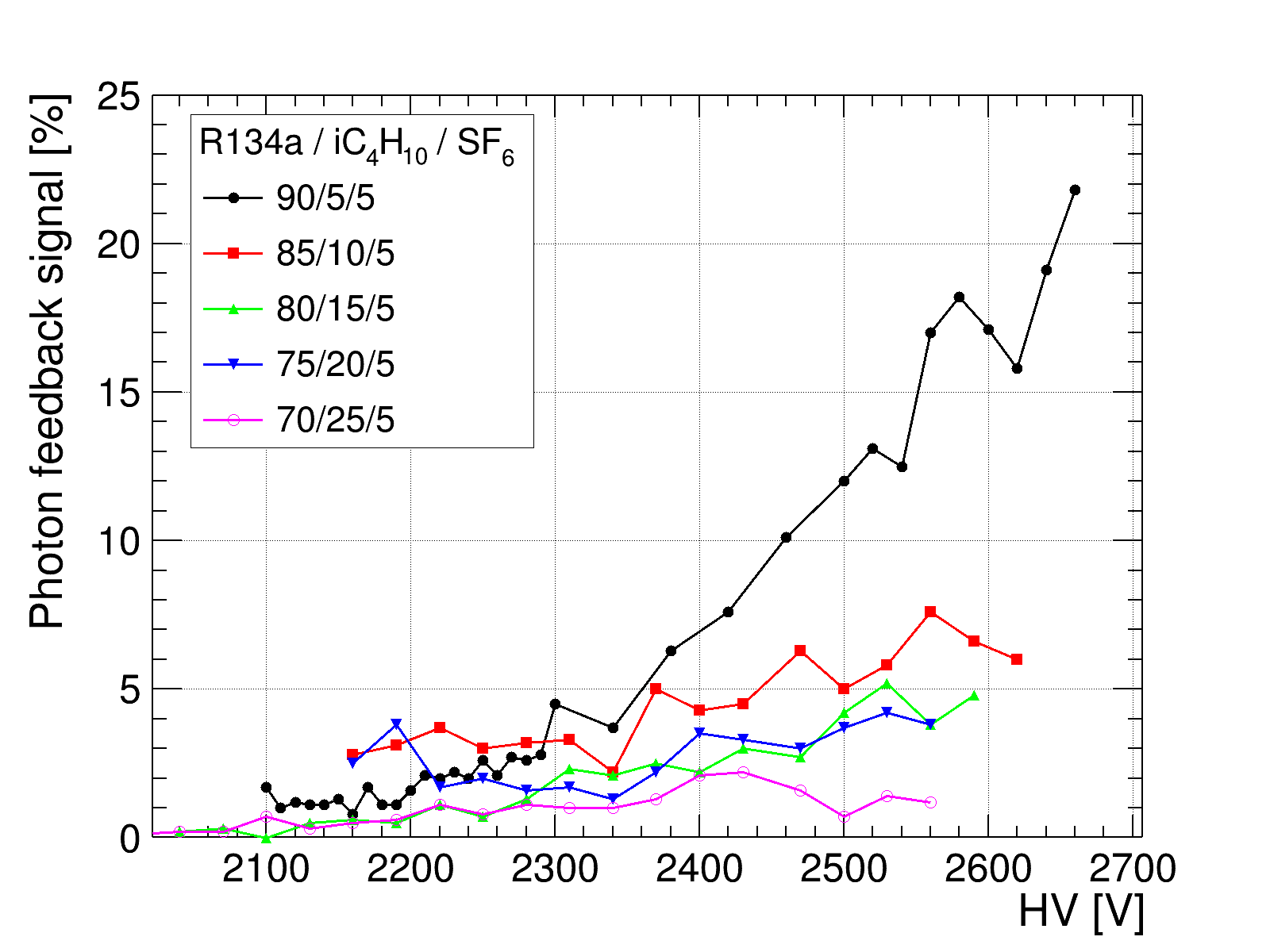}
    \caption{The ratio of signals with photon feedback at different gas ratios.}
    \label{feedback}
\end{figure}

Since $iC_{4}H_{10}$ has a strong absorption of UV photons, the fraction of $iC_{4}H_{10}$ was tuned to study this effect, noting that the ratio of $SF_{6}$ was fixed at 5\%. The ratio of signal with photon feedback (signals like Fig. \ref{twoE}) at different $iC_{4}H_{110}$ fraction is shown in Fig. \ref{feedback}. It is clear that at high voltage, with an increase in the $iC_{4}H_{10}$ iC4H10 fraction, the probability of photon feedback is significantly reduced, indicating that increasing the proportion of $iC_{4}H_{10}$ increases the absorption of UV photons.

\begin{figure}
    \centering
    \subfloat[]{\label{QRatio}\includegraphics[width=0.5\linewidth]{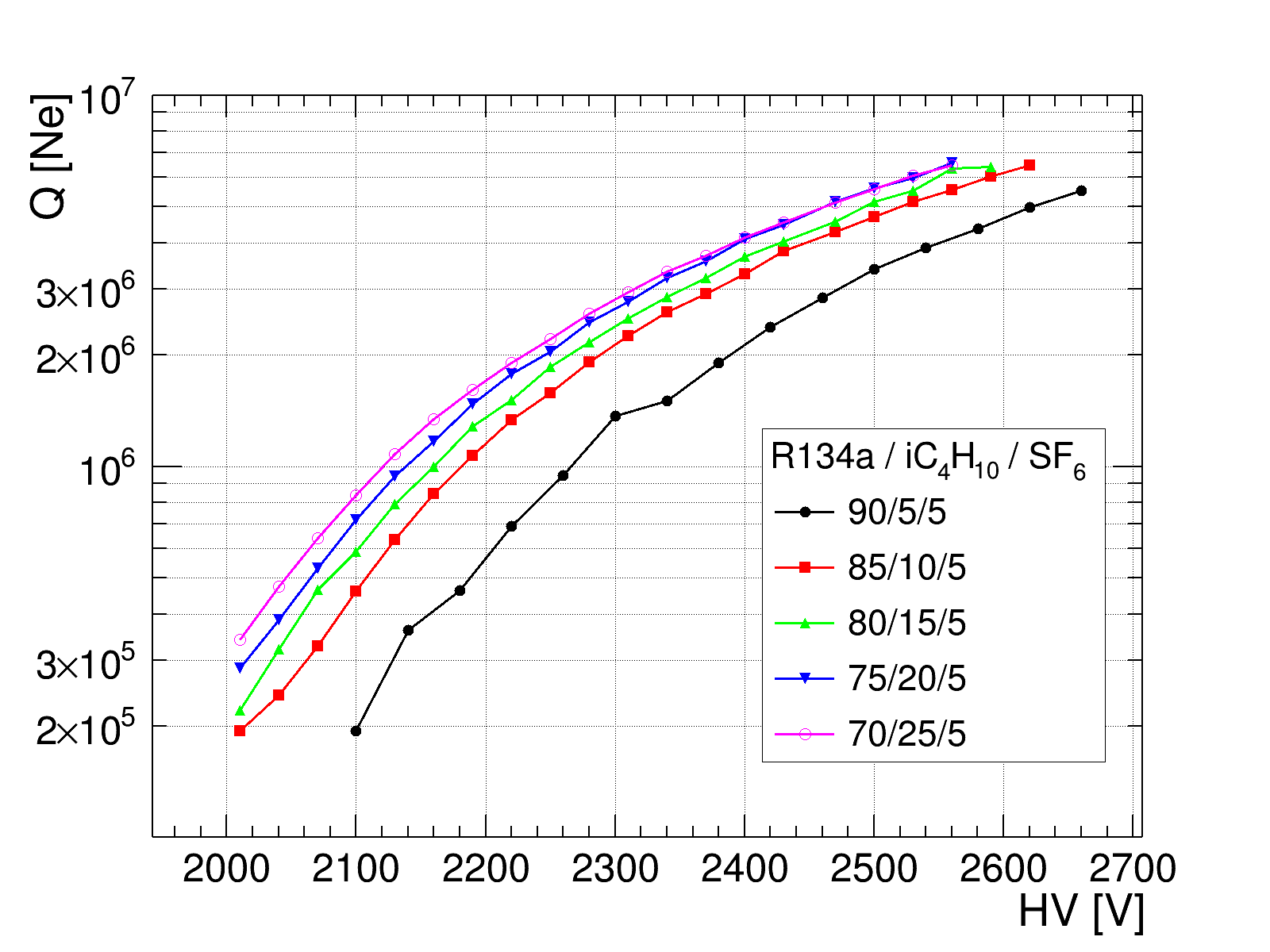}}
    \subfloat[]{\label{TRatio}\includegraphics[width=0.5\linewidth]{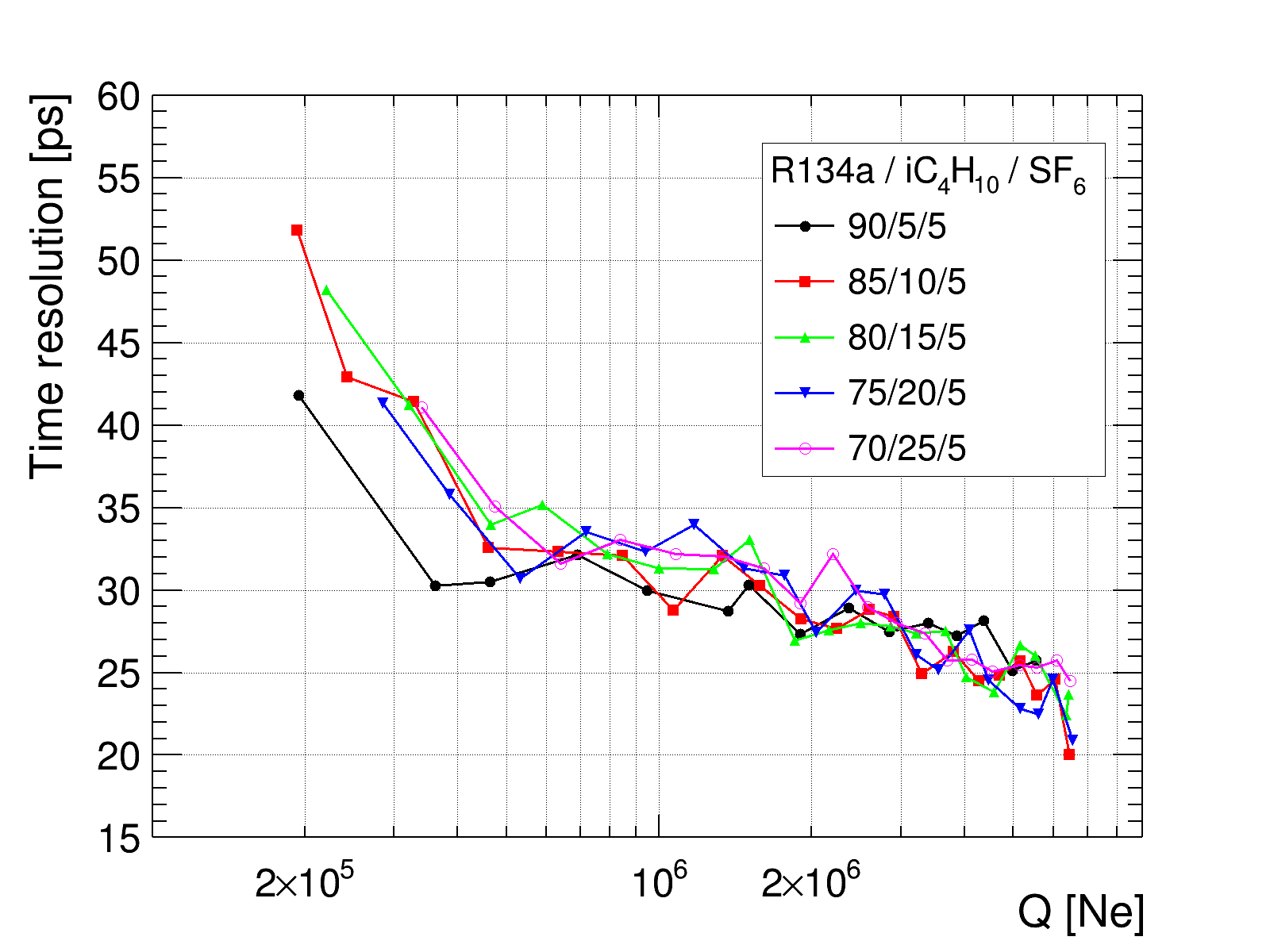}}
    
    \caption{The signal charges Q and time resolution of single photoelectron at different $iC_{4}H_{10}$ fractions.}
    \label{DifRatio}
\end{figure}

As the fraction of $iC_{4}H_{10}$ increases (thereby the fraction of R134a decreases), the maximum HV free of streamer gradually decreases, as shown in Fig. \ref{QR134a}. For example, in the $R134a/iC_{4}H_{10}/SF_{6}$ ratio of 90/5/5, there is no streamer until 2750V, but at 70/25/5, the streamer becomes visible at 2600V. The gain (signal charge) and time resolution are also shown in Fig. \ref{DifRatio}. The gain at the same voltage increases with an increase in the fraction of $iC_{4}H_{10}$. No significant change was observed in the time resolution as a function of the signal charge. It is worth noting that the single photoelectron time resolution of 20.3 ps was obtained with 85/10/5 ratio, at HV = 2620 V.

The results of the MRPC gas show that the amplitude of a single photoelectron signal is large, with a signal charge Q of $10^{5}\sim 10^{7}$ Ne. The rise time ($\sim$ 0.45 ns) indicates that the electron drift velocity is large. These features lead to a very good time resolution. With different gas ratios, the time resolution of single photoelectron is about 20 $\sim$ 52 ps. Increasing the proportion of $iC_{4}H_{10}$ (reducing R134a) can increase the absorption of UV photons, thus reduce the probability of photon feedback, meanwhile the gain at the voltage increases, but the maximum working voltage will decreases.

\subsection{Summary of test}

\begin{figure}
    \centering
    \includegraphics[width=1\linewidth]{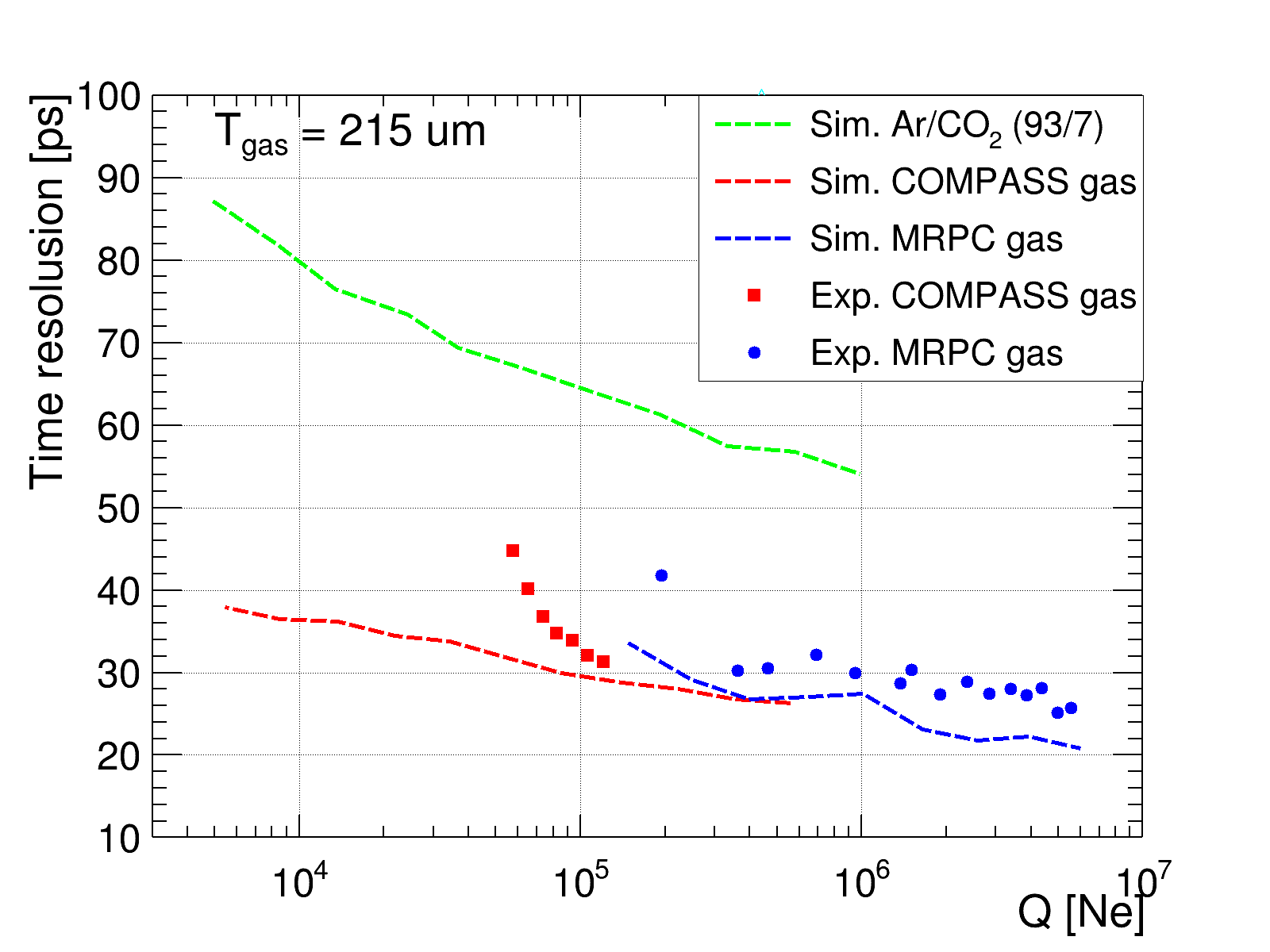}
    \caption{Comparison of laser test results with Garfield++ simulation.}
    \label{Compare}
\end{figure}

The laser test results in different gases are compared with the Garfield++ simulation, as shown in Fig. \ref{Compare}. The test results of COMPASS gas and MRPC gas are in good agreement with the simulation, especially at high gain, where good S/N ratio was achieved. The simulated performance with $Ar/CO_{2}$(93/7) gas mixture is also shown, but the test result is not compared due to poor S/N ratio in the experiment.


The working voltage, signal amplitude (with 26 dB), rise time, signal width (FWHM), signal charge and time resolution in different gases are summarized in Table \ref{TestSummary}. The best timing performance are obtained in MRPC gas, reaching a single photoelectron time resolution of 20.3 ps. Increasing the proportion of $iC_{4}H_{10}$ can significantly reduce the occurrence of photon feedback while the time resolution has no significant change. For COMPASS gas, single photoelectron time resolution of $\sim$ 31.3 ps can be achieved, with much lower gain ($\sim 10^{5}$) compared to MRPC gas, which may be beneficial for high-rate performance.


\begin{table}[]
\caption{Summary of single photoelectron test in diffrent gas.}
\label{TestSummary}
\centering
\resizebox{\linewidth}{!}{
   \begin{tabular}{|c|c|c|c|c|c|c|}
        \hline
       Gas & Working HV & amplitude& Rise time & FWHM & Q& Time resolution \\
         &[V] & (26 dB) [mV] & [ns] &[ns] & [Ne] &[ps] \\
       \hline
       $Ar/CO_{2}(93/7)$ & 775$\sim$850 & 15$\sim$20 & 1.28$\sim$1.35 & 2.6$\sim$2.9 & $2\cdot 10^{5}\sim 5\cdot 10^{5}$ & 90$\sim$120 \\
       \hline
       $Ne/CF_{4}/C_{2}H_{6}(80/10/10)$ & 780$\sim$840 & 25$\sim$38 & 0.36$\sim$0.39 & 0.55$\sim$0.59 & $6\cdot 10^{4}\sim 2\cdot 10^{5}$ & 31$\sim$45 \\
       \hline
       $R134a/iC_{4}H_{10}/SF_{6}$ & 2100$\sim$2700 & 30$\sim$650 & 0.41$\sim$0.49 & 0.94$\sim$1.1 & $2\cdot 10^{5}\sim 7\cdot  10^{6}$ & 20$\sim$52 \\
       \hline
   \end{tabular}
    \label{tab:my_label}
}
\end{table}

\section{Conclusion}

In order to meet the needs of high rate and high time resolution in future high energy physics experiments, we try to develop a gaseous photodetector. Through the simulation of Garfield++, it is found that the time resolution of a single photoelectron can be better than 30 ps with appropriate gas and gap thickness. When the initial electron is launched at a fixed point, the position fluctuation of the primary electron is eliminated, so the photoelectric RPC can obtain better time resolution than the typical RPC. We designed and developed a prototype of gaseous photodetector with RPC structure. The performance of the detector is extensively tested by an ultraviolet laser.

The rate capability of the resistive detector is closely related to the resistivity of material. We have constructed and tested the photoelectric RPC prototype with a variety of float glasses with different resistance. The results show that the low resistance floating glass ($\sim1.4\cdot10^{10}$ $\Omega\cdot cm$) significantly improves the capability while maintaining excellent timing resolution.


The detector was tested in different gases, the best results were obtained in MRPC gas, the single photoelectron time resolution best reachs 20.3 ps at a gain of $6\cdot10^{6}$. Increasing the proportion of $iC_{4}H_{10}$ (reducing R134a) can reduce the probability of photon feedback while barely changing the time resolution and maximum gain.

The prototype of the gaseous photodetector with RPC structure developed in this paper can achieve a high rate and very good single photoelectron time resolution in MRPC gas, but its disadvantage is that the ion back flow (IBF) is 1, which will affect the life of the photocathode, so a robust photocathode (such as DLC) is needed.

In addition, since the position of the initial photoelectron and the path length of the avalanche are both fixed, the characteristics of the single photoelectron signal in different gases can be analyzed in a controlled manner. Therefore, this kind of photodetector can be applied to study the properties of different gases, e.g. electron drift velocity and gas gain can be deduced from measured signal charge and rise time.
\section*{Acknowledgments}
The authors thank the USTC high-energy physics group. This project is supported by the National Natural Science Foundation of China under grant No. U11927901,12205296 and 11975228, the State Key Laboratory of Particle Detection and Electronics under grant No. SKLPDE-ZZ-202416 and the USTC Research Funds of the Double First-Class Initiative under Grant No.WK2030000052.  Special thank goes to Professor Qing Luo from National Synchrotron Radiation Laboratory (China) for help with FEM calculation (ANSYS).


\bibliography{mybibfile}

\begin{thebibliography}{10}
\expandafter\ifx\csname url\endcsname\relax
  \def\url#1{\texttt{#1}}\fi
\expandafter\ifx\csname urlprefix\endcsname\relax\def\urlprefix{URL }\fi
\expandafter\ifx\csname href\endcsname\relax
  \def\href#1#2{#2} \def\path#1{#1}\fi

\bibitem{firstRPC1}
R.~Santonico, R.~Cardarelli, Development of resistive plate counters, Nucl. Instrum. Methods Phys. Res. 187~(2-3) (1981) 377--380.
\newblock \href {http://dx.doi.org/10.1016/0029-554X(81)90363-3} {\path{doi:10.1016/0029-554X(81)90363-3}}.

\bibitem{firstRPC2}
R.~Santonico, R.~Cardarelli, et~al., Progress in resistive plate counters, Nucl. Instrum. Methods Phys. Res. A 263~(1) (1988) 20--25.
\newblock \href {http://dx.doi.org/10.1016/0168-9002(88)91011-X} {\path{doi:10.1016/0168-9002(88)91011-X}}.

\bibitem{triggerRPCinATLAS}
M.~Alviggi, et~al., {Cosmic ray test station for ATLAS RPC}, Nucl. Instrum. Methods Phys. Res. A 508~(1-2) (2003) 124--127, 6th International Workshop on Resistive Plate Chambers and Related Detectors, COIMBRA, PORTUGAL, NOV 26-27, 2001.
\newblock \href {http://dx.doi.org/10.1016/S0168-9002(03)01291-9} {\path{doi:10.1016/S0168-9002(03)01291-9}}.

\bibitem{timingRPC}
P.~Fonte, {High-resolution timing of MIPs with RPCs - a model}, Nucl. Instrum. Methods Phys. Res. A 456~(1-2) (2000) 6--10, 5th International Workshop on Resistive Plate Chambers and Related Detectors, BARI, ITALY, OCT 28-29, 1999.
\newblock \href {http://dx.doi.org/10.1016/S0168-9002(00)00953-0} {\path{doi:10.1016/S0168-9002(00)00953-0}}.

\bibitem{timingRPCforTOF}
P.~Fonte, R.~Marques, et~al., {High-resolution RPCs for large TOF system}, Nucl. Instrum. Methods Phys. Res. A 449~(1-2) (2000) 295--301.
\newblock \href {http://dx.doi.org/10.1016/S0168-9002(99)01299-1} {\path{doi:10.1016/S0168-9002(99)01299-1}}.

\bibitem{firstMRPC}
E.~{Cerron Zeballos}, et~al., \href{https://www.sciencedirect.com/science/article/pii/0168900296001581}{{A new type of resistive plate chamber: The multigap RPC}}, Nucl. Instrum. Methods Phys. Res. A 374~(1) (1996) 132--135.
\newblock \href {http://dx.doi.org/https://doi.org/10.1016/0168-9002(96)00158-1} {\path{doi:https://doi.org/10.1016/0168-9002(96)00158-1}}.
\newline\urlprefix\url{https://www.sciencedirect.com/science/article/pii/0168900296001581}

\bibitem{MRPCTOF}
Akindinov, et~al., The multigap resistive plate chamber as a time-of-flight detector, Nucl. Instrum. Methods Phys. Res. A 456~(1-2) (2000) 16--22, 5th International Workshop on Resistive Plate Chambers and Related Detectors, BARI, ITALY, OCT 28-29, 1999.
\newblock \href {http://dx.doi.org/10.1016/S0168-9002(00)00954-2} {\path{doi:10.1016/S0168-9002(00)00954-2}}.

\bibitem{rateofRPC}
M.~Abbrescia, New studies on the rate capability of resistive gaseous detectors, Journal of Physics (2022) 012155 (4 pp.)\href {http://dx.doi.org/10.1088/1742-6596/2374/1/012155} {\path{doi:10.1088/1742-6596/2374/1/012155}}.

\bibitem{HighRateModeofRPC}
Carboni, et~al., {A model for RPC detectors operating at high rate}, Nucl. Instrum. Methods Phys. Res. A 498~(1-3) (2003) 135--142.
\newblock \href {http://dx.doi.org/10.1016/S0168-9002(02)02082-X} {\path{doi:10.1016/S0168-9002(02)02082-X}}.

\bibitem{LowRMRPC}
G.~Garillot, et~al., {Operation of a low resistivity glass MRPC at high rate using ecological gas}, Nucl. Instrum. Methods Phys. Res. A 1061.
\newblock \href {http://dx.doi.org/10.1016/j.nima.2024.169104} {\path{doi:10.1016/j.nima.2024.169104}}.

\bibitem{Solid20ps}
Arrington, et~al., {The solenoidal large intensity device (SoLID) for JLab 12 GeV}, {Journal of Physics G: Nuclear and Particle Physics} 50~(11).
\newblock \href {http://dx.doi.org/10.1088/1361-6471/acda21} {\path{doi:10.1088/1361-6471/acda21}}.

\bibitem{photoE_applyinGasdetector}
P.~Mine, Photoemissive materials and their application to gaseous detectors, Nucl. Instrum. Methods Phys. Res. A 343~(1) (1994) 99--108, 1st Workshop on Ring Imaging Cherenkov Detectors, BARI, ITALY, JUN 02-05, 1993.
\newblock \href {http://dx.doi.org/10.1016/0168-9002(94)90538-X} {\path{doi:10.1016/0168-9002(94)90538-X}}.

\bibitem{pad-photonDetector-Cherenkev}
R.~Arnold, et~al., A fast-cathode pad-photon detector for cherenkov ring imaging, Nucl. Instrum. Methods Phys. Res. A 314~(3) (1992) 465--494.
\newblock \href {http://dx.doi.org/10.1016/0168-9002(92)90239-Z} {\path{doi:10.1016/0168-9002(92)90239-Z}}.

\bibitem{PhotoGasDecAndApplication}
T.~Francke, V.~Peskov, Photosensitive gaseous detectors and their applications, Nucl. Instrum. Methods Phys. Res. A 525~(1-2) (2004) 1--5, international Conference on Imaging Techniques in Subatomic Physics, Astrophysics, Medicine, Biology and Industry, Stockholm, SWEDEN, JUN 24-27, 2003.
\newblock \href {http://dx.doi.org/10.1016/j.nima.2004.03.017} {\path{doi:10.1016/j.nima.2004.03.017}}.

\bibitem{PICOSEC}
{PICOSEC-Micromegas Collaboration}, Single photoelectron time resolution studies of the picosec-micromegas detector, Journal of Instrumentation 15~(4), 15th Topical Seminar on Innovative Particle and Radiation Detectors, Siena, ITALY, OCT 14-17, 2019.
\newblock \href {http://dx.doi.org/10.1088/1748-0221/15/04/C04053} {\path{doi:10.1088/1748-0221/15/04/C04053}}.

\bibitem{PPCwithCsI}
G.~Charpak, et~al., {Investigation of operation of parallel-plate avalanche chamber with a CsI photocathode under high-gain conditions}, Nucl. Instrum. Methods Phys. Res. A 307~(1) (1991) 63--68.
\newblock \href {http://dx.doi.org/10.1016/0168-9002(91)90131-9} {\path{doi:10.1016/0168-9002(91)90131-9}}.

\bibitem{photoRPCinRICH}
P.~Carlson, et~al., {Beyond the RICH: innovative photosensitive gaseous detectors for new fields of applications}, Nucl. Instrum. Methods Phys. Res. A 502~(1) (2003) 189--194, 4th International Workshop on Ring Imaging Cherenkov Detectors (RICH 2002), PYLOS, GREECE, JUN 05-10, 2002.
\newblock \href {http://dx.doi.org/10.1016/S0168-9002(03)00272-9} {\path{doi:10.1016/S0168-9002(03)00272-9}}.

\bibitem{hignRatePositionPhotoRPC}
T.~Francke, et~al., {High rate, high position resolution photosensitive RPCs}, Nucl. Instrum. Methods Phys. Res. A 533~(1-2) (2004) 163--168, 7th International Workshop on resistive Plate Chambers and Related Detectors, Clermont Ferrand, FRANCE, OCT 20-22, 2003.
\newblock \href {http://dx.doi.org/10.1016/j.nima.2004.07.021} {\path{doi:10.1016/j.nima.2004.07.021}}.

\bibitem{RPCwithCsI}
P.~Fonte, et~al., {Novel single photon detectors for UV imaging}, Nucl. Instrum. Methods Phys. Res. A 553~(1-2) (2005) 30--34, 5th International Workshop on Ring Imaging Cherenkov Detectors (RICH2004), Playa de Carmen, MEXICO, NOV 30-DEC 05, 2004.
\newblock \href {http://dx.doi.org/10.1016/j.nima.2005.08.002} {\path{doi:10.1016/j.nima.2005.08.002}}.

\bibitem{photoRPC25ps}
K.~Matsuoka, R.~Okubo, Y.~Adachi, Demonstration of a 25-picosecond single-photon time resolution with gaseous photomultiplication, Nucl. Instrum. Methods Phys. Res. A 1053.
\newblock \href {http://dx.doi.org/10.1016/j.nima.2023.168378} {\path{doi:10.1016/j.nima.2023.168378}}.

\bibitem{Raether1964ElectronAA}
H.~Raether, \href{https://api.semanticscholar.org/CorpusID:92932558}{Electron avalanches and breakdown in gases}, 1964.
\newline\urlprefix\url{https://api.semanticscholar.org/CorpusID:92932558}

\bibitem{induceSignalofRPC}
W.~Riegler, Induced signals in resistive plate chambers, Nucl. Instrum. Methods Phys. Res. A 491~(1-2) (2002) 258--271.
\newblock \href {http://dx.doi.org/10.1016/S0168-9002(02)01169-5} {\path{doi:10.1016/S0168-9002(02)01169-5}}.

\bibitem{MagboltzMC}
S.~Biagi, Monte carlo simulation of electron drift and diffusion in counting gases under the influence of electric and magnetic fields, Nucl. Instrum. Methods Phys. Res. A 421~(1-2) (1999) 234--240.
\newblock \href {http://dx.doi.org/10.1016/S0168-9002(98)01233-9} {\path{doi:10.1016/S0168-9002(98)01233-9}}.

\bibitem{magboltzWeb}
\href{https://ref.web.cern.ch/ref/CERN/CNL/2000/001/magboltz}{[link]}.
\newline\urlprefix\url{https://ref.web.cern.ch/ref/CERN/CNL/2000/001/magboltz}

\bibitem{garfieldWeb}
\href{https://garfieldpp.web.cern.ch/garfieldpp/}{[link]}.
\newline\urlprefix\url{https://garfieldpp.web.cern.ch/garfieldpp/}

\bibitem{RPC2003}
W.~Riegler, C.~Lippmann, R.~Veenhof, Detector physics and simulation of resistive plate chambers, Nucl. Instrum. Methods Phys. Res. A 500~(1-3) (2003) 144--162.
\newblock \href {http://dx.doi.org/10.1016/S0168-9002(03)00337-1} {\path{doi:10.1016/S0168-9002(03)00337-1}}.

\bibitem{BookRGD_testResitivity}
{Materials and Aging in Resistive Plate Chambers}, John Wiley \& Sons, Ltd, 2018, Ch.~6, pp. 223--225.
\newblock \href {http://dx.doi.org/https://doi.org/10.1002/9783527698691.ch6} {\path{doi:https://doi.org/10.1002/9783527698691.ch6}}.

\bibitem{spaceCharge}
C.~Lippmann, W.~Riegler, B.~Schnizer, Space charge effects and induced signals in resistive plate chambers, Nucl. Instrum. Methods Phys. Res. A 508~(1-2) (2003) 19--22, 6th International Workshop on Resistive Plate Chambers and Related Detectors, COIMBRA, PORTUGAL, NOV 26-27, 2001.
\newblock \href {http://dx.doi.org/10.1016/S0168-9002(03)01270-1} {\path{doi:10.1016/S0168-9002(03)01270-1}}.

\end{thebibliography}


\end{document}